\documentclass[aps,prd,reprint,superscriptaddress,floatfix]{revtex4-2}
\usepackage[T1]{fontenc}
\usepackage{lmodern}
\usepackage{amsmath}
\usepackage{amsfonts}
\usepackage{bm}
\usepackage{graphicx}
\usepackage{microtype}
\usepackage[hidelinks]{hyperref}

\newcommand{\Var}{\operatorname{Var}}
\newcommand{\Cov}{\operatorname{Cov}}

\begin{document}
\setcounter{secnumdepth}{2}

\title{Flexible Proposal Distributions for Gravitational-Wave Population Inference with Gaussian Mixture Models}

\author{Simon M. C. Yeung}
\affiliation{Leonard E. Parker Center for Gravitation, Cosmology, and Astrophysics,
University of Wisconsin--Milwaukee, Milwaukee, WI 53201, USA}
\author{Ignacio Maga\~na Hernandez}
\affiliation{McWilliams Center for Cosmology and Astrophysics, Department of Physics, Carnegie Mellon University, Pittsburgh, PA 15213, USA}
\author{Jolien Creighton}
\affiliation{Leonard E. Parker Center for Gravitation, Cosmology, and Astrophysics,
University of Wisconsin--Milwaukee, Milwaukee, WI 53201, USA}

\begin{abstract}
Population analyses of gravitational-wave (GW) sources use hierarchical Bayesian inference to constrain astrophysical models while accounting for selection effects.
The likelihood is usually estimated by Monte Carlo integration over finite sets of event posterior samples and simulated injections.
As catalogs grow, more samples are needed to control the Monte Carlo variance, but the available event samples are fixed by parameter estimation.
Sharp population features can also lie in regions poorly sampled by the event posteriors.
We fit Gaussian mixture models (GMMs) to the event posteriors and evaluate the likelihood by importance sampling with two proposals: an ensemble of event posteriors and a mixture of that ensemble with a fiducial population.
Together with a GMM representation of detected injections, this allows additional samples to be drawn for both event and selection integrals.
Increasing the sample count reduces Monte Carlo variance but leaves uncertainty from fitting the densities to finite training sets.
In simulated catalogs, the GMM-based estimators show no detectable additional bias relative to the posterior-sample estimator; their calibration is comparable within the limitations of the tests.
An application to GWTC-3 with a different mass model gives consistent results across the estimators.
The method separates the numerical sampling budget from the original sample count, while requiring explicit checks of density-fit uncertainty and approximation error.
\end{abstract}

\maketitle

\section{Introduction}\label{sec:introduction}

Since the first direct detection of gravitational waves (GWs) in 2015 \cite{150914}, increasing detector sensitivity and observing time of the Advanced LIGO \cite{Aasi_2015}, Advanced Virgo \cite{Acernese_2015}, and KAGRA \cite{Akutsu_2021} have expanded the catalog of GW events. 
The latest Gravitational-Wave Transient Catalog (GWTC-5.0) of the LIGO-Virgo-KAGRA (LVK) Collaboration contains 390 cumulative compact-binary candidates with astrophysical probability $p_{\rm astro}\geq 0.5$ \cite{GWTC5-intro, GWTC5ObsandUpdates}.
This growing catalog has enabled population studies of compact binaries \cite{GWTC1pop, GWTC2pop, GWTC3pop, GWTC4pop, GWTC5pop, Fishbach_2018, Fishbach_2017, Callister_2021, Farr_2017, Roulet_2019, Ray_2023, Callister_2024, Wysocki_2019}.

Population inference uses ensembles of GW events to constrain astrophysical models describing the masses, spins, and redshift distribution of compact binaries, providing insights across astrophysics and cosmology \cite{Loredo_2004, Mandel_2019, Thrane2019_BayesianGW, Vitale_2021}.
For binary neutron star (BNS) systems, joint analyses of the mass distribution and equation of state constrain neutron star matter \cite{Abbott2018_EOS, Chatziioannou_2020, Golomb_EOS}.
For binary black hole (BBH) systems, the mass spectrum features, the spin distributions, and redshift evolution probe formation channels and cosmic evolution \cite{Fishbach_2018, Mandel_2022, Zevin_2021}.
In addition, gravitational-wave observations constrain cosmology, with or without electromagnetic counterparts \cite{Abbott2017_H0, Ezquiaga_2022}.

These properties are inferred within a hierarchical Bayesian framework that account for observational selection effects \cite{Mandel_2019, Thrane2019_BayesianGW, Vitale_2021}.
Within this framework, the population is described by parametrized distributions over the source parameters, characterized by a set of hyperparameters.
For example, the BBH primary-mass distribution is modeled as a broken power law with two Gaussian peaks, whose hyperparameters include two power-law slopes, the break mass, the minimum and maximum masses, and the location, width, and the weights of the peaks \cite{GWTC4pop, GWTC5pop}.
Spin magnitudes are modeled as truncated-Gaussian distributions, and spin tilts as a mixture of an isotropic component and truncated Gaussians, with hyperparameters for the means, widths, and the mixing fraction  \cite{Vitale_2022, GWTC5pop}.
The merger rate is modeled as a power law in $(1+z)$ with slope $\gamma$ \cite{Fishbach_2018}.

The likelihood used to infer these hyperparameters involves per-event integrals over source parameters and an integral for the selection function, both estimated as Monte Carlo sums \cite{Mandel_2019, essick2022}.
For comparable, independent event contributions, the Monte Carlo standard deviation of the catalog log-likelihood scales approximately as $\sqrt{N_{\rm obs}/N_{\rm samp}}$, where $N_{\rm obs}$ is the number of events and $N_{\rm samp}$ is the number of samples per event \cite{Farr_2019, Talbot_pain, essick2022, heinzel2025}.
Maintaining this error level therefore requires the per-event sample count to grow with the catalog.
Current population analyses instead rely on finite sets of posterior samples from single-event parameter estimation (PE) \cite{gwpop1, Talbot_2019}.
Their size is limited by the cost of waveform generation and likelihood evaluation \cite{purrer2019}, and analyses may also downsample events to a common Monte Carlo budget \cite{Abbott2023_GWTC3}.
Additional PE samples require further computation, so a fixed sample set can become a source of increasing likelihood uncertainty as the catalog grows \cite{Talbot_pain, heinzel2025}.

The same holds for selection function, the Monte Carlo uncertainty already excludes a substantial part of the hyperparameter space for spin distributions with the injection set of the third observing run \cite{Talbot_sel}.
The issue will worsen as the catalog grows, with the upgraded detectors projected to detect of order $10^3$ BBH events by the end of the fifth observing run \cite{Abbott_2020_LRR, Kiendrebeogo_2023, farah2024}.
Future third-generation (3G) detectors are expected to increase the number of observed binaries by orders of magnitude \cite{CE, ET}, with forecasts of $10^{5}$--$10^{6}$ BBH detections per year \cite{Borhanian_2024}.
Thus, there is a present and increasing need to control the Monte Carlo uncertainty in population inference.

One approach is density estimation, which approximates a continuous probability density from a finite sample set \cite{Silverman_1986}.
Examples include kernel density estimation, GMMs, and normalizing flows \cite{Kobyzev_2021, nf1, Silverman_1986, Bishop_2006_PRML, maf}.
A fitted density can be evaluated throughout its modeled support and sampled repeatedly.
This separates the Monte Carlo sampling budget from the number of training samples, although it does not remove uncertainty or bias in the fitted density.
It also permits importance-sampling proposals tailored to sharp population features \cite{Golomb_EOS}.

Exact physical relations, such as an equation of state or a cosmological distance--redshift relation, can restrict the population to narrow regions or lower-dimensional surfaces in parameter space \cite{wysocki2020, Landry_2021, Golomb_EOS, Ezquiaga_2022}.
A finite PE sample set may poorly resolve those regions even when the posterior density there is nonzero, making direct reweighting unreliable.
This sampling limitation differs from genuinely negligible posterior support, which contributes negligibly to the likelihood integral.
We focus here on consistency and error control; a full treatment of exact physical relations is left to future work.

GMMs have already been used in GW population inference.
Talbot and Thrane \cite{Talbot_sel} fitted detected injections to estimate selection effects, obtaining smaller statistical errors than direct injection reweighting at comparable cost.
Golomb and Talbot \cite{Golomb_EOS} fitted BNS posteriors to infer the mass distribution and equation of state jointly.
Here we use GMM event posteriors with shared proposal distributions, including a population-informed mixture that concentrates samples in regions favored by a fiducial population.
Unlike methods that learn the population distribution itself \cite{RuheNormalizingStudy}, or emulate the hyperparameter posterior \cite{Mould_2025}, our approach retains an explicitly specified population model and changes the numerical evaluation of its likelihood \cite{Mandel_2019, Thrane2019_BayesianGW}.
We test the estimators with a toy model, selection-efficiency calculations, likelihood-variance estimates, and repeated simulated catalogs.
We then compare them on GWTC-3 \cite{GWTC3pop}, a fixed catalog used here as a controlled benchmark rather than an analysis of the latest detections.

Section~\ref{sec:methods} defines the population models and estimators.
Section~\ref{sec:results} presents the validation tests and GWTC-3 comparison, and Sec.~\ref{sec:discussion} discusses limitations and extensions.
The smoothing functions, variance derivation, and density-fit diagnostics are given in Appendices~\ref{app:smoothing}--\ref{app:fit}.

\section{Methods}\label{sec:methods}

We write the source-frame parameters as $\vartheta=\{m_1,m_2,z\}$ and their detector-frame counterparts as $\theta=\{m_{1,\rm det},m_{2,\rm det},d_L\}$.
Here $m_1\geq m_2$ are the component masses, $z$ is redshift, and $d_L$ is luminosity distance.
Population models are defined on the source-frame parameters.

Both population models use the same power-law redshift evolution \cite{Fishbach_2018}:
\begin{equation}
    \begin{aligned}
    p(z|\gamma) &\;=\; \frac{\mathcal{C}_z(\gamma)^{-1}}{(1+z)}\,\frac{dV_c}{dz}\,(1+z)^{\gamma} \\
    \mathcal{C}_z(\gamma) &\;=\; \int_0^{z_{\max}}\frac{dV_c}{dz}\,(1+z)^{\gamma-1}\,dz,
    \end{aligned}
    \label{eq:pz}
\end{equation}
Here $\mathcal{C}_z(\gamma)$ normalizes the distribution over $0\leq z\leq 2$, $dV_c/dz$ is the differential comoving volume in the Planck 2015 cosmology, and the factor $(1+z)^{-1}$ converts source-frame time to detector-frame time \cite{planck}.

We use different mass models for the mock data challenge (MDC) and the GWTC-3 analysis.
Both contain a power law plus a Gaussian peak (PLP), but differ in the pairing of component masses and the treatment of mass cutoffs.
For the MDC, we adopt the sharp-edged mass model of Ray et al.\ \cite{ray2024_mdc}:
\begin{equation}
    p(m_1,m_2|\Lambda) =
    p_{\rm PLP}(m_1|\Lambda_m)
    p_q(q|m_1, \beta)
    \frac{1}{m_1},
    \label{eq:conditional_plp}
\end{equation}
Here $q=m_2/m_1$, and $1/m_1$ is the Jacobian for changing variables from $q$ to $m_2$. The component distributions are
\begin{align}
    p_{\rm PLP}(m_1|\Lambda_{\rm m})
    &= (1-\lambda)\,\frac{(1-\alpha)m_1^{-\alpha}}
    {m_{\max}^{1-\alpha}-m_{\min}^{1-\alpha}} \notag\\
    &\quad +\lambda\,\frac{e^{-\frac12[(m_1-\mu)/\sigma]^2}}
    {\sqrt{2\pi}\,\sigma},
    \label{eq:plp_sharp}\\
    p_q(q|m_1,\beta)
    &=\frac{(1+\beta)q^\beta}
    {1-(m_{\min}/m_1)^{1+\beta}}.
    \label{eq:pq_sharp}
\end{align}
The mass support is $m_{\min}\leq m_2\leq m_1\leq m_{\max}$, with zero density outside.
$\Lambda_{\rm m}=\{\alpha,m_{\min},m_{\max},\beta,\mu,\sigma,\lambda\}$ denotes the mass-model hyperparameters.
The power-law component has hard cutoffs at $m_{\min}$ and $m_{\max}$; at $\alpha=1$, its normalization is understood as the logarithmic limit.

For GWTC-3, we use a symmetric pairing model \cite{farah2024, farah2024things, Pairing}: both component masses follow a common distribution $p_{\rm PLP}^S$, with their correlation described by a mass-ratio pairing function.
\begin{equation}
    \begin{split}
    p(m_1,m_2|\Lambda)\propto{}&p_{\rm PLP}^S(m_1|\Lambda_{\rm m}')\,
    p_{\rm PLP}^S(m_2|\Lambda_{\rm m}')\\
    &\times f(m_1,m_2|\beta_{\rm pair}),
    \end{split}
    \label{eq:pairing}
\end{equation}
where the smoothed mass distribution is
\begin{equation}
\begin{split}
    p^{S}_{\rm PLP}(m|\Lambda_{\rm m}')
    ={}&(1-\lambda)\,
    \frac{S_{\rm lo}(m)\,S_{\rm hi}(m)\,m^{-\alpha}}
    {\mathcal{C}_{\rm m}(\Lambda_{\rm m}')}\\
    &+\lambda\,
    \frac{e^{-\frac12[(m-\mu)/\sigma]^2}}
    {\sqrt{2\pi}\,\sigma}.
\end{split}
\label{eq:plp_smooth}
\end{equation}
\begin{equation}
    \mathcal{C}_{\rm m}(\Lambda_{\rm m}')
    = \int S_{\rm lo}(m)\,S_{\rm hi}(m)\,m^{-\alpha}\,dm .
    \label{eq:plp_normalization}
\end{equation}
The pairing function is
\begin{equation}
    f(m_1, m_2 | \beta_{\rm pair}) = \left(\frac{m_2}{m_1}\right)^{\beta_{\rm pair}}.
    \label{eq:pairing_q}
\end{equation}
The functions $S_{\rm lo}$ and $S_{\rm hi}$ smooth the lower and upper cutoffs of the power-law component; $\mathcal{C}_{\rm m}$ is its normalization.
Their explicit forms are given in Appendix~\ref{app:smoothing}.
The MDC model has eight hyperparameters:
\begin{equation}
\Lambda=\{\gamma,\alpha,m_{\min},m_{\max},\beta,\mu,\sigma,\lambda\}.
\end{equation}
The GWTC-3 model replaces $\beta$ with $\beta_{\rm pair}$ and adds the two smoothing widths $\delta m_{\min}$ and $\delta m_{\max}$, giving ten hyperparameters, denoted by $\Lambda'$.

The exponents in Eqs.~\eqref{eq:pq_sharp} and \eqref{eq:pairing_q} describe different quantities.
In the MDC model, $\beta$ is the mass-ratio slope.
In the symmetric model, changing variables from $m_2$ to $q$ at fixed $m_1$ gives
\begin{equation}
    p(q|m_1)\propto q^{\beta_{\rm pair}}\,p^S_{\rm PLP}(m_1q|\Lambda_{\rm m}'),
\end{equation}
so the secondary mass retains the same underlying mass function as the primary.
Away from the Gaussian peak and the smoothed edges, the power-law component gives $p(q|m_1)\propto q^{\beta_{\rm pair}-\alpha}$.
Its effective mass-ratio slope in that regime is therefore $\beta_{\rm pair}-\alpha$, not $\beta_{\rm pair}$.
Their recovered values are therefore not directly comparable.

We infer the hyperparameters with a likelihood that accounts for the observed data $\{d_i\}$ and the detector network's selection effects.
Table~\ref{tab:priors} gives the priors; $\beta$ and $\beta_{\rm pair}$ have the same prior range despite their different meanings.

\begin{table}[ht]
\centering
\caption{
    Uniform priors for the population hyperparameters. Smoothing widths apply only to the GWTC-3 model; the MDC and GWTC-3 models use $\beta$ and $\beta_{\rm pair}$, respectively.
}
\label{tab:priors}
\renewcommand{\arraystretch}{1.2}
\setlength{\tabcolsep}{8pt}
\begin{tabular}{|l|c|}
\hline\hline
\textbf{Hyperparameter} & \textbf{Prior} \\
\hline
$\alpha$        & $\mathcal{U}(0,\ 6)$ \\
\hline
$m_{\min}$      & $\mathcal{U}(2,\ 10)\,M_\odot$ \\
\hline
$m_{\max}$      & $\mathcal{U}(50,\ 100)\,M_\odot$ \\
\hline
$\delta m_{\min}$ & $\mathcal{U}(1,\ 100)\,M_\odot$ \\
\hline
$\delta m_{\max}$ & $\mathcal{U}(1,\ 100)\,M_\odot$ \\ 
\hline
$\mu$           & $\mathcal{U}(20,\ 50)\,M_\odot$ \\
\hline
$\sigma$        & $\mathcal{U}(1,\ 10)\,M_\odot$ \\
\hline
$\lambda$       & $\mathcal{U}(0,\ 1)$ \\
\hline
$\beta$, $\beta_{\rm pair}$         & $\mathcal{U}(0,\ 6)$ \\
\hline
$\gamma$        & $\mathcal{U}(0,\ 10)$ \\
\hline\hline
\end{tabular}
\end{table}

Conditional on the observed event count, the population likelihood is \cite{Mandel_2019, Thrane2019_BayesianGW, Abbott2023_GWTC3}

\begin{equation}
    \mathcal{L}(\{d_i\}|\Lambda) \propto 
    \xi(\Lambda)^{-N_{\mathrm{obs}}}
    \prod_{i=1}^{N_{\mathrm{obs}}} 
    \int d\theta \mathcal{L}_i(d_i | \theta) p(\theta|\Lambda)
    ,
\label{eq:total_likelihood}
\end{equation}
where $\mathcal{L}_i(d_i|\theta)$ is the per-event likelihood and $\xi(\Lambda)$ is the detection efficiency.
The detection efficiency is the fraction of the population detectable by the network:

\begin{equation}
    \xi(\Lambda)=\int d\vartheta\,P_{\rm det}\bigl(\theta(\vartheta)\bigr)\,p_{\rm pop}(\vartheta|\Lambda).
    \label{eq:selection}
\end{equation}
Here $P_{\rm det}(\theta)$ is the probability that an event with detector-frame parameters $\theta$ is detected.
This integral is numerically evaluated using a set of $N_{\mathrm{inj}}$ samples drawn from a mock population $p_{\mathrm{inj}}$. The efficiency is thus approximated as the weighted mean of successful injections:

\begin{equation}
    \hat\xi(\Lambda)=\frac{1}{N_{\rm inj}}\sum_{j=1}^{N_{\rm inj}}\frac{p_{\rm pop}(\vartheta_j|\Lambda)}{p_{\rm inj}(\vartheta_j|\Lambda_{\rm inj})}\,I_j,
    \label{eq:standard_selection}
\end{equation}
where $N_{\rm inj}$ is the total number of generated injections and $I_j$ indicates whether injection $j$ passes the detection criterion. In the mock-data campaign, $I_j=1$ for a recovered signal-to-noise ratio (SNR) greater than 9 and is zero otherwise.

A commonly used diagnostic requires an effective sample size $N_{\rm eff}=(\sum_l w_l)^2/\sum_l w_l^2>4N_{\rm obs}$, where $w_l$ are the Monte Carlo weights \cite{essick2022, Farr_2019}.
For direct injection reweighting, $N_{\rm eff}$ is bounded by the number of available injections, whereas the required precision grows with catalog size.
A GMM allows more evaluation samples to be drawn from a fitted density.
This can raise the conditional Monte Carlo effective sample size, but it neither adds independent injection information nor removes density-fit uncertainty.

A GMM represents a density $p(\theta)$ as a weighted sum of $K$ multivariate Gaussian components \cite{Bishop_2006_PRML}:
\begin{equation}
    p(\theta) \approx \mathcal{D}(\theta) = 
    \sum_{k=1}^K a_{k} \mathcal{N}(\theta|\mu_{k}, \Sigma_{k}),
\end{equation}
Here $a_k\geq0$ and $\sum_k a_k=1$, and $\mathcal N(\theta|\mu_k,\Sigma_k)$ is a multivariate Gaussian with mean $\mu_k$ and covariance $\Sigma_k$.

Following Talbot and Thrane \cite{Talbot_sel}, we fit the density of detected injections in transformed coordinates.
We start from $m_{1,\rm det}$, $\operatorname{logit}(q)$, and $d_L$, and map each coordinate to an unbounded space using its empirical cumulative distribution and a probit transformation.
Let $T:\theta\mapsto\theta'$ denote the full transformation, including any whitening, and let $\mathcal D(\theta')$ be the fitted GMM.
The detection probability is then approximated by
\begin{equation}
    \widehat P_{\rm det}(\theta)
    =\eta\,\frac{\mathcal D(T(\theta))}{p_{\rm inj}(\theta|\Lambda_{\rm inj})}
    \left|\det\frac{\partial T}{\partial\theta}\right|,
    \label{eq:detection_density}
\end{equation}
where $\eta$ is the injection population's detection efficiency, defined below.
The Jacobian converts the normalized density of detected injections back to detector-frame coordinates; the factor $\eta$ restores its absolute normalization.

We then evaluate the selection efficiency $\xi(\Lambda)$ with importance sampling.

\begin{align}
    \hat\xi(\Lambda)
    &=\frac{1}{N_{\rm GMM}}\sum_{j=1}^{N_{\rm GMM}}
    \frac{P_{\rm det}(\theta(\vartheta_j))\,p_{\rm pop}(\vartheta_j|\Lambda)}
    {q_{\rm sel}(\vartheta_j)}\notag\\
    &=\frac{\eta}{N_{\rm GMM}}\sum_{j=1}^{N_{\rm GMM}}
    \frac{p_{\rm pop}(\vartheta_j|\Lambda)}
    {p_{\rm inj}(\vartheta_j|\Lambda_{\rm inj})},
    \label{eq:selection_gmm}
\end{align}
with samples $\vartheta_j\sim q_{\rm sel}$, where
\begin{equation}
    q_{\rm sel}(\vartheta)=\frac{P_{\rm det}(\theta(\vartheta))\,
    p_{\rm inj}(\vartheta|\Lambda_{\rm inj})}{\eta}
    \label{eq:sel_proposal}
\end{equation}
and
\begin{equation}
    \eta=\int d\vartheta\,P_{\rm det}(\theta(\vartheta))\,
    p_{\rm inj}(\vartheta|\Lambda_{\rm inj}).
\end{equation}
All densities in this importance ratio use source-frame coordinates.
In practice, we approximate $q_{\rm sel}$ with the fitted detected-injection density and estimate $\eta$ by the detected fraction of the injection campaign.
Equation~\eqref{eq:selection_gmm} is therefore conditional on that approximation.
Concentrating samples in the detectable region can improve efficiency relative to drawing directly from the full population.

The population-marginalized likelihood for event $i$ is
\begin{equation}
    \mathcal{L}_i(\Lambda) = \int d\theta\, \mathcal{L}(d_i|\theta)\, p(\theta|\Lambda).
\label{eq:single_event_def}
\end{equation}
Here
\begin{equation}
    p(\theta|\Lambda) 
    = p_{\rm pop}(\vartheta(\theta)|\Lambda)
    \left|\frac{\partial \vartheta}{\partial \theta}\right|
\label{eq:source_to_det}
\end{equation}
and 
\begin{equation}
    \left|\frac{\partial \vartheta}{\partial \theta}\right|
    = (1+z)^{-2}\left(\frac{\mathrm{d}d_L}{\mathrm{d}z}\right)^{-1}
\label{eq:jacobian}
\end{equation}
The determinant in Eq.~\eqref{eq:jacobian} converts the density from source-frame to detector-frame coordinates.

A standard calculation in LVK population analyses applies Bayes' theorem to Eq.~\eqref{eq:single_event_def} \cite{Thrane2019_BayesianGW, Vitale_2021, Abbott2023_GWTC3}.
For a PE prior $\pi(\theta)$, posterior $p(\theta|d_i)$, and evidence $Z_i=\int d\theta\,\mathcal L(d_i|\theta)\pi(\theta)$, the estimator is
\begin{equation}
    \widehat{\mathcal L}_i(\Lambda) =
    \frac{Z_i}{N_{\mathrm{PE}}} 
    \sum_{j=1}^{N_{\mathrm{PE}}}
    \frac{p(\theta_j|\Lambda)}{\pi(\theta_j)},
\label{eq:standard_mc}
\end{equation}
where $\theta_j\sim p(\theta|d_i)$ and $N_{\rm PE}$ is the number of posterior samples.
We use the PE prior associated with the samples; the distance dependence assumed here is $\pi(\theta)\propto d_L^2$ within its stated support.

For independent samples with finite weight variance, the Monte Carlo variance scales inversely with sample count.
The available PE sample count depends on waveform cost, sampler settings, thinning, autocorrelation, and convergence \cite{bilby, dynesty, binning}.
Reweighting a fixed sample set cannot increase that count.
A GMM instead provides a continuous approximation from which further evaluation samples can be drawn, reducing Monte Carlo variance conditional on the fit.

One option is to change the proposal in Eq.~\eqref{eq:standard_mc}: draw from the population and evaluate the fitted event-posterior densities,

\begin{equation}
    \widehat{\mathcal L}_i(\Lambda) =
    \frac{Z_i}{N_{\mathrm{POP}}} 
    \sum_{j=1}^{N_{\mathrm{POP}}}
    \frac{\hat{p}(\theta_j|d_i)}{\pi(\theta_j)},
    \label{eq:population-drawn}
\end{equation}
where $\theta_j\sim p(\theta|\Lambda)$ and $\hat p(\theta|d_i)$ is a GMM fitted to the PE samples.
This proposal can resolve narrow population features, but many population draws may have negligible likelihood for a given event.
The proposal also changes with $\Lambda$, so samples and GMM densities must be recomputed at each likelihood evaluation.

\begin{table*}[t]
\centering
\caption{Per-event likelihood estimators and proposal distributions, all expressed in detector-frame coordinates $\theta$.
Weights omit the common evidence factor $Z_i$.
The PE prior is $\pi(\theta)$; the assumed distance dependence is $d_L^2$.
Per-event GMMs $\hat p_i\equiv\hat p(\theta|d_i)$ use full-covariance components selected by the Bayesian information criterion (BIC).
The ensemble is $q_{\rm E}=N_{\rm obs}^{-1}\sum_i\hat p_i$.
For Mixture, $f=0.1$ and $\Lambda_0$ is the median from an initial population-inference run using PE.
The cost column gives the dependence on sample and event counts after density fitting and any caching; it omits the dependence on GMM component count.
The analyses use $N_{\rm obs}=69$ events.}
\label{tab:methods}
\renewcommand{\arraystretch}{1.65}
\setlength{\tabcolsep}{5pt}
\small
\begin{tabular}{lllll}
\hline\hline
Method & Proposal $q(\theta)$ & Per-event weight & Proposal update & Per-call cost\\
\hline
PE & $p(\theta|d_i)$ & $\dfrac{p(\theta_j|\Lambda)}{\pi(\theta_j)}$
& Fixed PE samples & $\mathcal O(N_{\rm PE}N_{\rm obs})$\\
POP & $p(\theta|\Lambda)$ & $\dfrac{\hat p(\theta_j|d_i)}{\pi(\theta_j)}$
& Redrawn at each $\Lambda$ & $\mathcal O(N_{\rm POP}N_{\rm obs})$\\
Ensemble & $q_{\rm E}(\theta)$ & $\dfrac{\hat p(\theta_j|d_i)p(\theta_j|\Lambda)}{\pi(\theta_j)q(\theta_j)}$
& Fixed and cached & $\mathcal O(N_{\rm GMM}N_{\rm obs})$\\
Mixture & $f q_{\rm E}(\theta)+(1-f)p(\theta|\Lambda_0)$ & $\dfrac{\hat p(\theta_j|d_i)p(\theta_j|\Lambda)}{\pi(\theta_j)q(\theta_j)}$
& Fixed at $\Lambda_0$ & $\mathcal O(N_{\rm GMM}N_{\rm obs})$\\
\hline\hline
\end{tabular}
\end{table*}

We use a shared importance-sampling proposal $q(\theta)$ to improve overlap with the regions favored by both the event posteriors and the population model:

\begin{equation}
    \widehat{\mathcal L}_i(\Lambda) =
    \frac{Z_i}{N_{\mathrm{GMM}}} 
    \sum_{j=1}^{N_{\mathrm{GMM}}}
    \frac{\hat{p}(\theta_j|d_i)p(\theta_j|\Lambda)} {\pi(\theta_j) q(\theta_j)}.
    \label{eq:importance_sampling}
\end{equation} 

Here $\theta_j\sim q(\theta)$, and the same sample pool is used for every event.
Our first proposal is the ensemble $q_{\rm E}(\theta)=N_{\rm obs}^{-1}\sum_i\hat p(\theta|d_i)$.
It concentrates samples in regions supported by the event posteriors, although it may still undersample a narrow region favored by the population.
To add population-informed draws, we use the mixture

\begin{equation}
    q(\theta) = 
    \frac{f}{N_{\mathrm{obs}}}
    \,\sum_{i=1}^{N_{\rm obs}} \hat{p}(\theta|d_i) + (1-f)\,p(\theta|\Lambda_0),
    \label{eq:mixture_proposal}
\end{equation}
where $f\in[0,1]$ is the fraction drawn from the ensemble and $\Lambda_0$ specifies a fixed reference population.
The limits $f=1$ and $f=0$ recover the ensemble and reference-population proposals, respectively.
A reference population with poor overlap with the likelihood integrand can produce large importance weights and a low effective sample size.
We choose $\Lambda_0$ from the posterior medians of an initial population-inference run using Eq.~\eqref{eq:standard_mc}.
Related strategies recycle earlier draws or combine importance-sampling proposals \cite{AMIS, owenzhou, Capp2004PopulationMC}.
We set $f=0.1$, drawing 10\% of the samples from the ensemble and 90\% from the reference population.
Unlike the population-drawn estimator, this proposal is fixed throughout the subsequent run, so the samples and GMM evaluations can be cached.

It is noted that a fitted GMM can extrapolate density into region where event has no or little posterior support.
We therefore apply a truncation to the event's GMM, restricting it to the 0.02th--99.98th percentile range of the training PE samples in each coordinate.
The truncation is applied to the event density in the numerator of Eq.(\ref{eq:importance_sampling}).
Proposal draws are not subjected to the truncation, as the draws are shared across the event's density estimates and any draw outside an event's range receives zero weight for that event.
This is close in spirit to the bounded GMM used in De Renzis et al. \cite{boundary_fix}.

We refer to Eq.~\eqref{eq:standard_mc} as \textbf{PE}, Eq.~\eqref{eq:population-drawn} as \textbf{POP}, Eq.~\eqref{eq:importance_sampling} with $q=q_{\rm E}$ as \textbf{Ensemble}, and the proposal in Eq.~\eqref{eq:mixture_proposal} as \textbf{Mixture}.
Table~\ref{tab:methods} summarizes the four estimators.

\section{Results}\label{sec:results}

\subsection{Toy-model consistency test}
We first compare the implementations with a two-dimensional Gaussian population in source-frame mass and redshift.
The means and standard deviations are $\mu_m=35\,M_\odot$, $\sigma_m=1\,M_\odot$, $\mu_z=1$, and $\sigma_z=0.1$.
We draw ten events, transform them to detector-frame coordinates, and assign each a bivariate normal posterior centered on its drawn $(m_{\rm det},d_L)$ values.
Each event has 1000 simulated posterior samples.
We fit a GMM to each event, select its component count by BIC, and use covariance regularization $\texttt{reg\_covar}=10^{-3}$ with a fixed random seed.
The population priors are $\mu_m\sim\mathcal U(25,45)\,M_\odot$, $\sigma_m\sim\mathcal U(0.1,10)\,M_\odot$, $\mu_z\sim\mathcal U(0.1,2)$, and $\sigma_z\sim\mathcal U(0.01,0.5)$.
We omit selection effects to isolate the event-likelihood calculation.
This test compares PE, POP, and Ensemble; Mixture uses the same importance-sampling expression as Ensemble with a different proposal.

\begin{figure}[tbp]
    \centering
    \includegraphics[width=\linewidth]{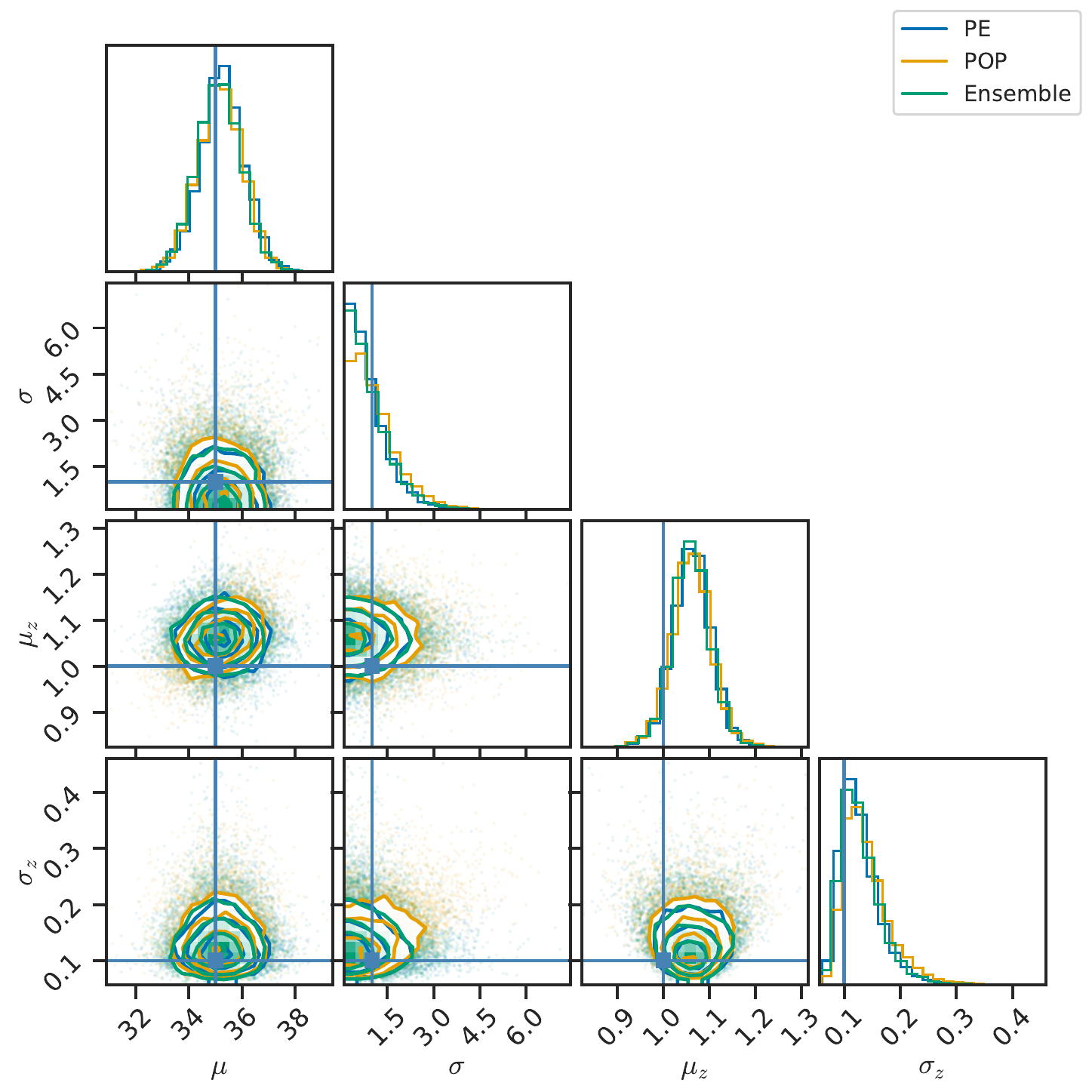}
    \caption{
    Hyperparameter posteriors for the Gaussian toy population, inferred with PE (blue), POP (orange), and Ensemble (green).
    The overlapping posteriors provide a check of consistency between the implementations.
    }
    \label{fig:proof_of_concept}
\end{figure}

Figure~\ref{fig:proof_of_concept} shows the recovered population means and standard deviations.
The means are localized near the injected values, and the three methods give substantially overlapping posteriors.
We find no detectable shift relative to PE in this test.

The estimators differ in computational cost despite their agreement in this test.
POP redraws samples and reevaluates the event densities whenever the hyperparameters change, and it can be inefficient when the population has little overlap with individual event posteriors.
We therefore restrict the remaining comparisons to PE, Ensemble, and Mixture.

\begin{figure*}[tp]
\centering
\begin{minipage}[t]{0.48\textwidth}
\centering
    \includegraphics[width=\linewidth]{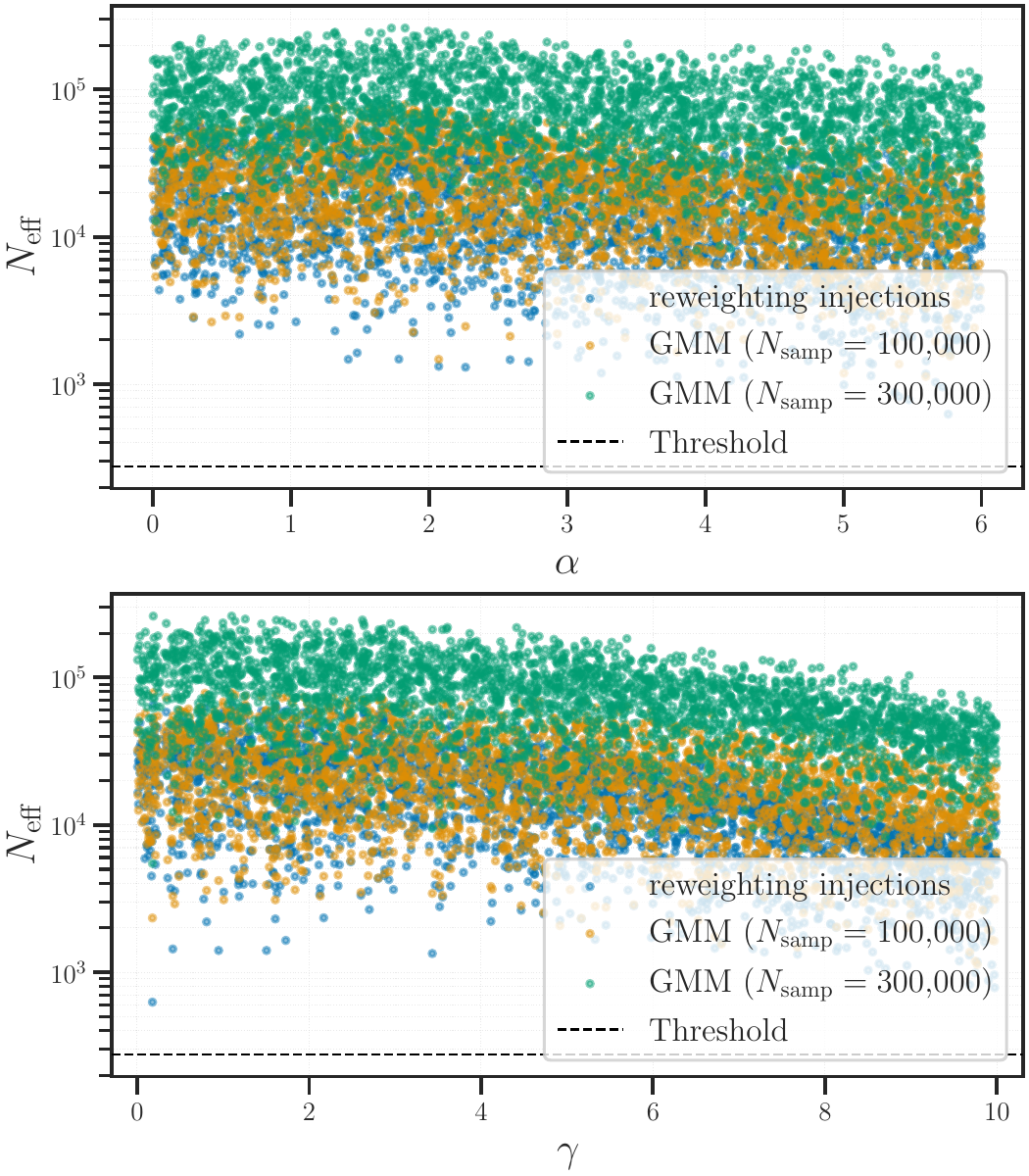}
    \caption{
    Effective sample size $N_{\rm eff}$ for the selection-efficiency estimator as a function of the mass power-law index $\alpha$ (top) and redshift exponent $\gamma$ (bottom).
    Direct injection reweighting is shown in blue; GMM draws use $N_{\rm samp}=100{,}000$ (orange) and $300{,}000$ (green).
    At the smaller sample count, the effective sample sizes are comparable to direct reweighting; increasing the GMM sample count raises the conditional Monte Carlo effective sample size.
    }
    \label{fig:Neff_plot}
\end{minipage}\hfill
\begin{minipage}[t]{0.48\textwidth}
\centering
    \includegraphics[width=\linewidth]{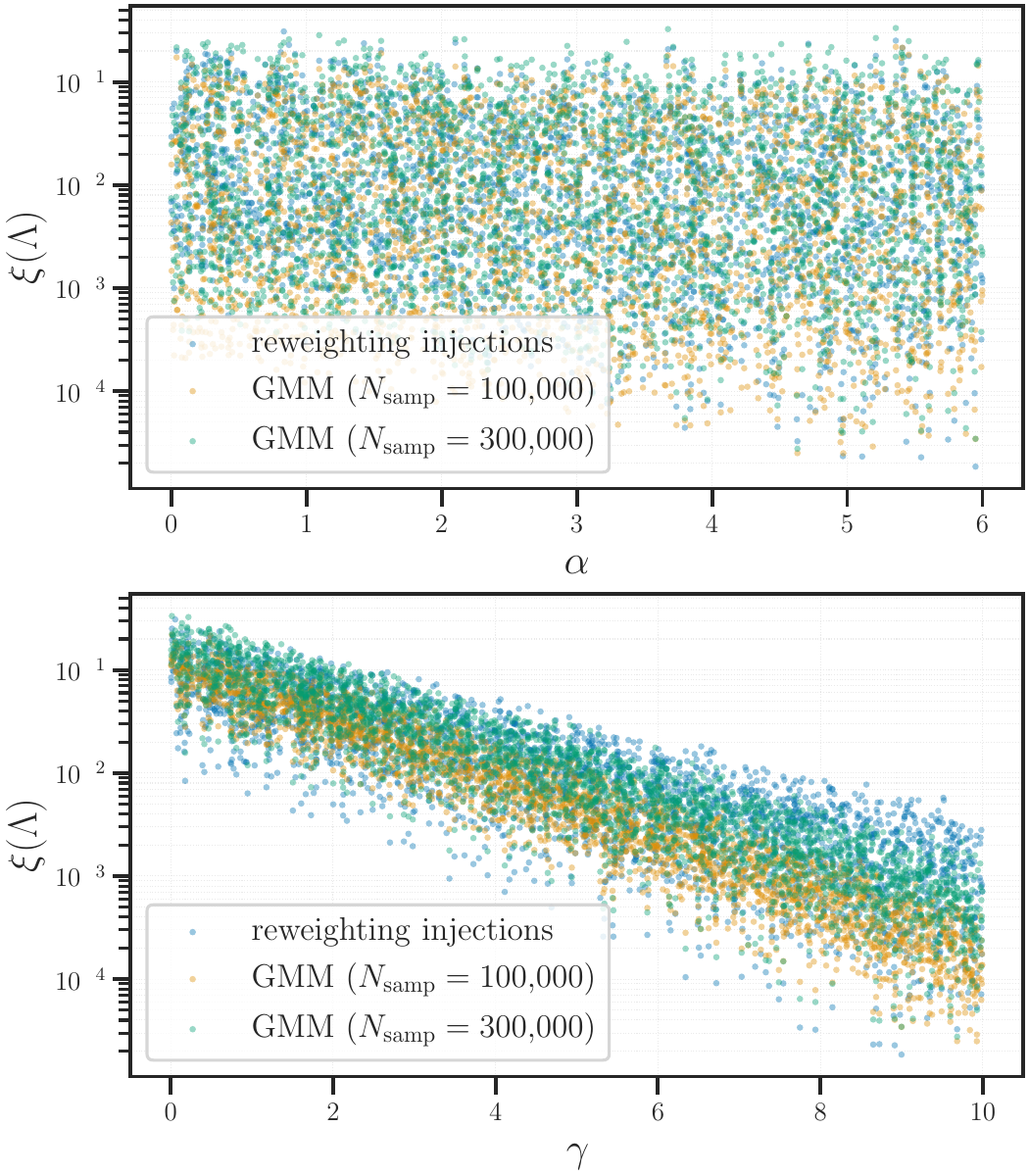}
    \caption{
    Selection efficiency $\xi(\Lambda)$ as a function of $\alpha$ (top) and $\gamma$ (bottom), with colors as in Fig.~\ref{fig:Neff_plot}.
    The plot shows 5000 hyperparameter evaluations.
    The two GMM sample counts give similar selection efficiencies, with no detectable shift as the sample count increases.
    }
    \label{fig:xi_plot}
\end{minipage}
\par\bigskip
\centering
\begin{minipage}[t]{0.48\textwidth}
\vspace{0pt}
\includegraphics[width=\linewidth]{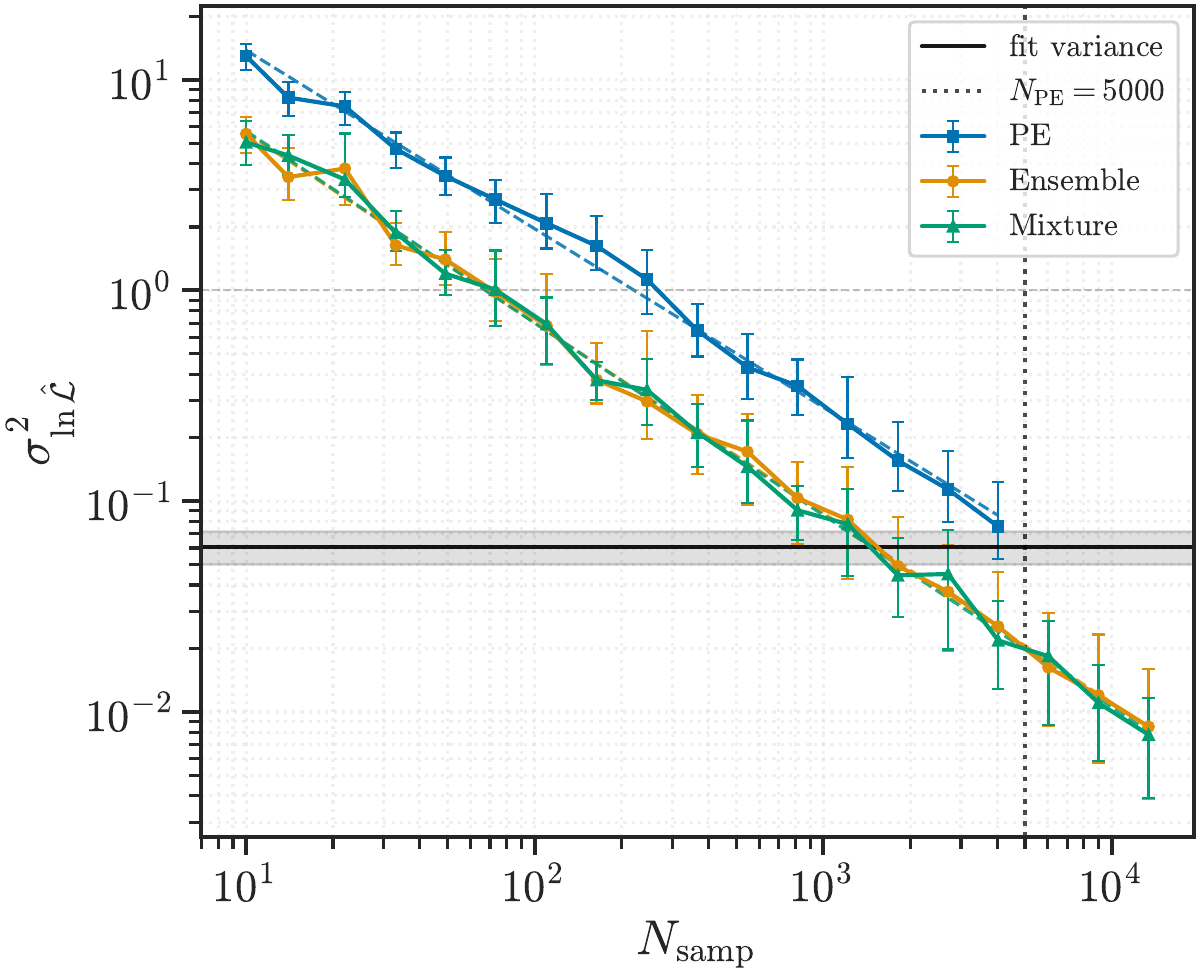}
\end{minipage}\hfill
\begin{minipage}[t]{0.48\textwidth}
\vspace{0pt}
\caption{
    Variance of the catalog sum of event log-likelihoods versus the per-event sample budget $N_{\rm samp}$ for PE (blue), Ensemble (orange), and Mixture (green).
    At each sample count, we evaluate the variance at 500 hyperparameter draws from the prior; markers and bars show the median and 16th--84th percentile range across those draws.
    The vertical dotted line marks the available PE sample count.
    The Monte Carlo variance decreases approximately as $1/N_{\rm samp}$; GMM-based draws allow the calculation to continue beyond the PE sample count.
    The horizontal black line and gray band show the median and $1\sigma$ band of the bootstrap fit-variance estimate.
    Its median is approximately $0.06$. The selection contribution is not included.
    }
    \label{fig:var}
\end{minipage}
\end{figure*}

\subsection{Mock catalogs and selection efficiency}
For the MDC, we draw BBHs from the population in Eqs.~\eqref{eq:pz} and \eqref{eq:conditional_plp}, with
\begin{align*}
    (\gamma,\alpha,\beta,\lambda)&=(3.0,4.0,1.5,0.04),\\
    (m_{\min},m_{\max},\mu,\sigma)&=(6.0,70.0,35.0,4.0)\,M_\odot.
\end{align*}
We transform the parameters to the detector frame and retain events with SNR greater than 9 using the Hanford and Livingston sensitivity curves from the third observing run (O3) \cite{Buikema_2020}.
We generate posterior samples with \texttt{GWMockCat} \cite{GWMockCat}.
A separate injection campaign samples a broad population using the same sensitivity curves and detection threshold to estimate $\xi(\Lambda)$.
The event pool contains 276 detections, from which we draw catalogs of 69 events.

We compare the selection estimators using the diagnostic $N_{\rm eff}>4N_{\rm obs}$ \cite{Farr_2019}.
We set $N_{\rm obs}=69$ to match the size of the GWTC-3 BBH population sample \cite{Abbott2023_GWTC3}; the mock SNR threshold is not the catalog's search-significance criterion.
Both estimators use the same recovered injections.
The standard estimator reweights them directly with Eq.~\eqref{eq:standard_selection}, whereas the GMM is fitted once and then sampled for Eq.~\eqref{eq:selection_gmm}.
We evaluate $N_{\rm eff}$ as a function of $\alpha$ and $\gamma$, drawing $100{,}000$ and $300{,}000$ GMM samples in separate calculations.

Figure~\ref{fig:Neff_plot} shows that all three calculations exceed $4N_{\rm obs}$ over the tested prior range.
At $N_{\rm samp}=100{,}000$, the GMM and direct-injection estimators have comparable effective sample sizes.
Increasing the GMM sample count to $300{,}000$ raises $N_{\rm eff}$, whereas direct reweighting is limited by the fixed injection set.
Figure~\ref{fig:xi_plot} compares the corresponding selection efficiencies.
The estimates agree closely as a function of $\alpha$ and substantially overlap as a function of $\gamma$, with no detectable additional bias relative to direct reweighting in this comparison.
These tests concern sampling precision conditional on the injection-density fit, not its accuracy outside the tested regions.

\subsection{Likelihood variance and density-fit uncertainty}
We next study the event contribution to the log-likelihood variance.
When the event and selection calculations use independent sample pools, the variance decomposes as \cite{Talbot_pain}
\begin{equation}
\begin{split}
\Var[\ln\widehat{\mathcal L}]={}&\Var\!\left[\sum_i\ln\widehat{\mathcal L}_i\right]\\
&+N_{\rm obs}^2\Var[\ln\hat\xi].
\end{split}
\label{eq:variance_summary}
\end{equation}
Large likelihood variance can bias population inference \cite{Talbot_pain, heinzel2025}.
We use $\Var[\ln\widehat{\mathcal L}]=1$ as a reference threshold \cite{Talbot_pain}, but report only the catalog sum of the event terms below, including their cross-covariances.
The selection contribution must also be included when assessing the total error budget.

Figure~\ref{fig:var} shows the event contribution to the variance, summarized over 500 prior draws at each sample budget.
All three estimators scale approximately as $1/N_{\rm samp}$.
PE is limited to the 5000 available posterior samples per event in this test; reducing its variance further would require more PE samples.
For Ensemble and Mixture, we draw up to $N_{\rm samp}=20{,}000$ from the fitted densities and reference population.
Both methods have lower variance than PE at matched sample counts, and their variances are comparable to one another.
The correlation structure differs: Ensemble and Mixture evaluate all event densities on a shared proposal-sample pool, producing cross-event Monte Carlo covariances.
PE instead uses independent posterior samples for each event.
Appendix~\ref{app:variance} derives the variance expressions, including these covariance terms.

Increasing the evaluation sample count does not remove uncertainty in a density fitted to finite PE samples.
We estimate this fit uncertainty by refitting each event's GMM to 50 bootstrap resamples of its posterior samples.
For each refit, we evaluate the event likelihood at 200 hyperparameter draws from the prior and measure the variance across refits.
We sum the per-event estimates under the independent-fit approximation described in Appendix~\ref{app:variance}; Eq.~\eqref{eq:fit_var} gives the estimator.
The median catalog fit variance is approximately $0.06$ in this test (Fig.~\ref{fig:var}).
It dominates once the conditional Monte Carlo variance falls below that level, so further evaluation draws offer little improvement in the combined variance.
This bootstrap estimate measures variability of the fitted densities, not their systematic approximation bias.

Here we downsample every event to a common PE sample count before fitting.
Using all available samples for each event may reduce fit uncertainty without rerunning PE, but the measured variance is not a proven upper bound.
Averaging densities fitted to bootstrap resamples is another possible variance-reduction strategy \cite{breiman1996bagging, ensembleGMM}; we do not test it here.

\subsection{Calibration across simulated catalogs}
For a calibrated posterior under repeated prior-predictive simulations, the posterior cumulative probability evaluated at the true parameter is uniformly distributed \cite{Cook2006, talts2020}.
Thus, for example, 30\% of true values should lie below the corresponding posterior 30th percentile.
We draw 20 hyperparameter sets from the priors in Table~\ref{tab:priors} and generate a pool of 276 detected events for each set.
From each pool, we construct eight catalogs of 69 events, sampling without replacement within each catalog, for 160 analyses per method.
Catalogs from the same pool can share events and are therefore correlated; this limits the interpretation of their aggregate calibration statistics.

\begin{figure*}[tp]
    \centering
    \includegraphics[width=\linewidth]{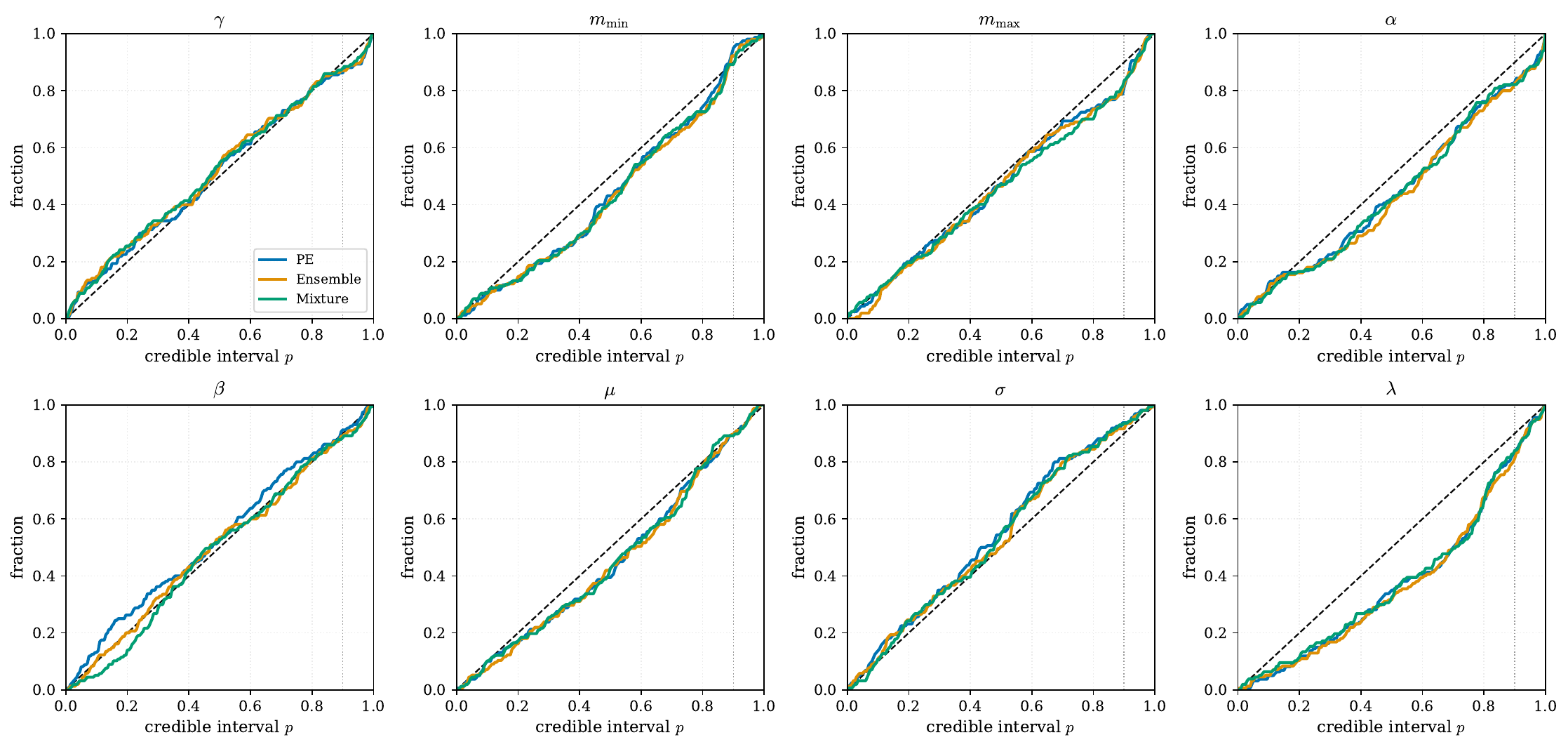}
    \caption{
    Probability--probability (PP) plots for the eight hyperparameters in 160 simulated catalogs.
    Each of 20 prior hyperparameter draws defines a pool of 276 detected events, from which eight catalogs of 69 events are sampled without replacement within each catalog; different catalogs may share events.
    The horizontal axis is a posterior quantile $p$, and the vertical axis is the fraction of catalogs whose true value lies at or below that quantile.
    A calibrated result follows the diagonal in the prior-predictive limit.
    The methods give similar curves, but $\lambda$ departs from the diagonal.
    Only four of the 20 injected $\lambda$ values exceed 0.5, limiting the calibration assessment for this parameter.
    }
    \label{fig:pp_plot}
\par\bigskip
\centering
    \includegraphics[width=\linewidth, keepaspectratio]{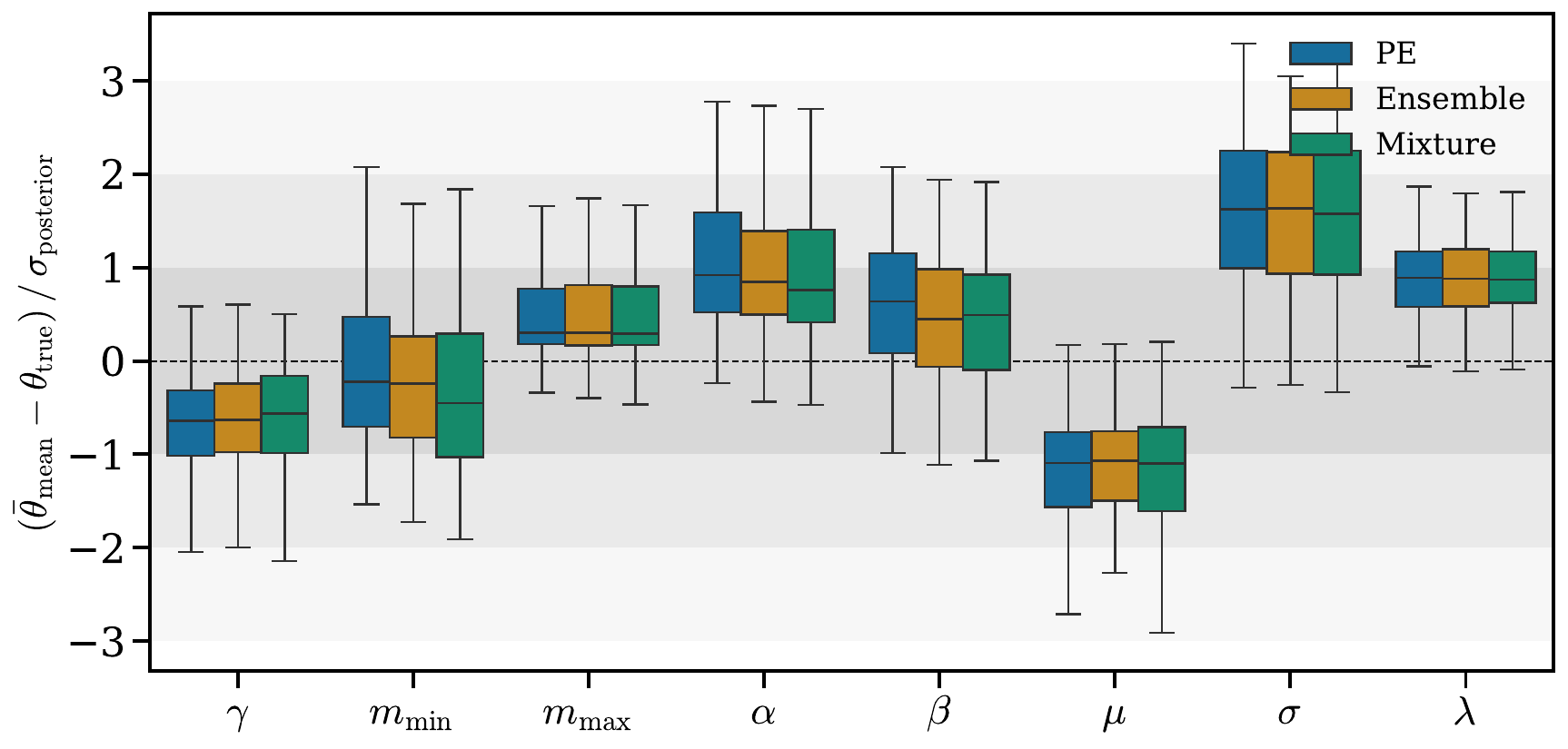}
    \caption{
    Pull distributions for the eight hyperparameters across 50 catalog realizations.
    Gray bands mark offsets within one, two, and three posterior standard deviations; the horizontal line marks zero offset.
    PE, Ensemble, and Mixture give overlapping distributions, with no detectable additional shift from the GMM estimators in this test.
    }
    \label{fig:pull_box}
\end{figure*}

Figure~\ref{fig:pp_plot} compares the three estimators.
Most parameters track the diagonal, whereas $\lambda$ shows a departure shared by all three methods.
Only four of the 20 injected $\lambda$ values exceed the prior mean of 0.5; this limited prior sampling, together with catalog overlap, prevents a firm attribution of the discrepancy.
The GMM-based curves are consistent with PE, but the test does not establish exact calibration for every parameter.

\subsection{Parameter recovery at a fixed population}
\begin{figure*}[tp]
    \centering
    \includegraphics[width=0.84\textwidth,height=0.70\textheight,keepaspectratio]{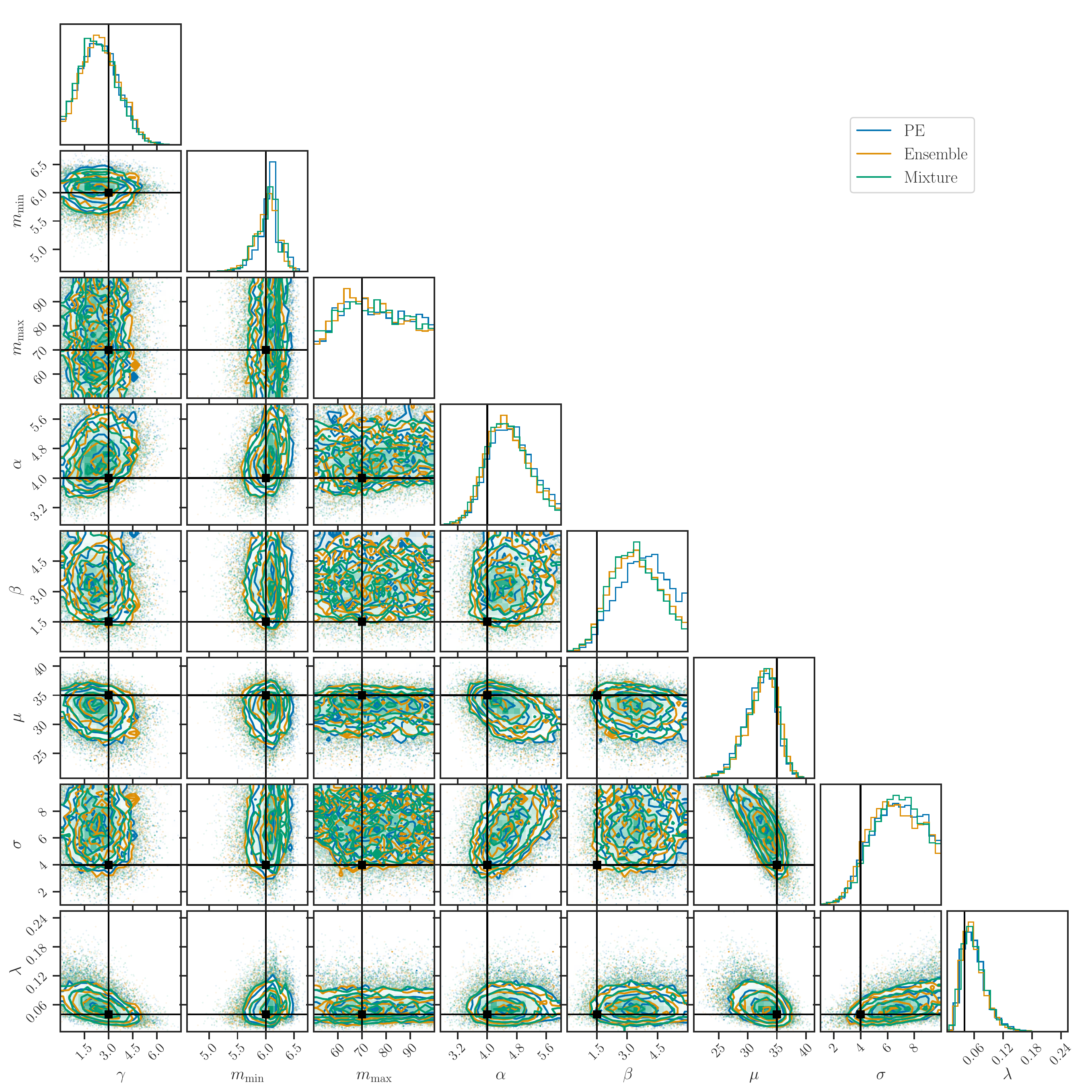}
    \caption{
    Hyperparameter posteriors for the sharp-edged MDC model in one catalog realization.
    PE (blue), Ensemble (orange), and Mixture (green) are compared with the injected values (black lines).
    The posteriors substantially overlap.
    }
    \label{fig:mdc_pos}
\par\medskip
\centering
    \includegraphics[width=0.48\textwidth]{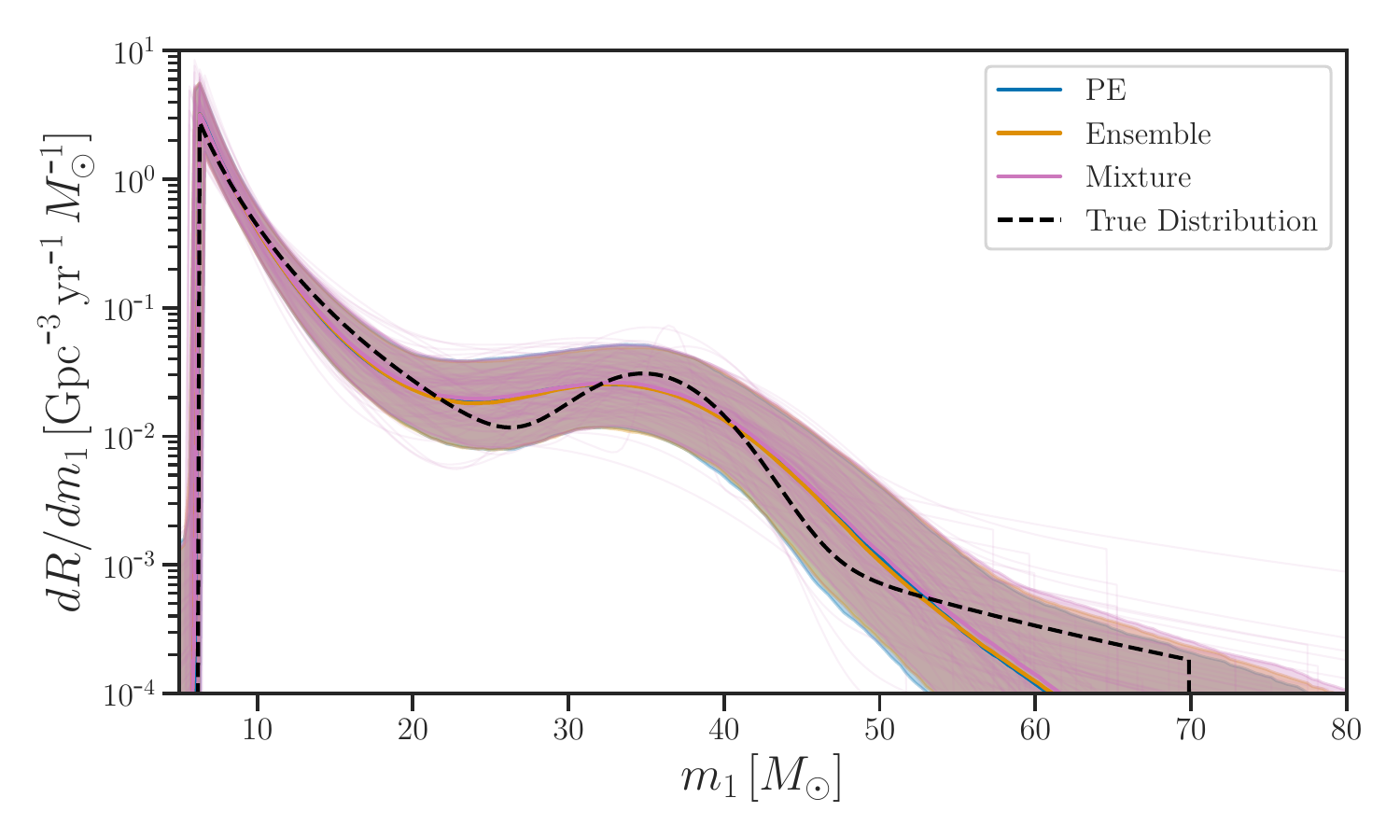}\hfill
    \includegraphics[width=0.48\textwidth]{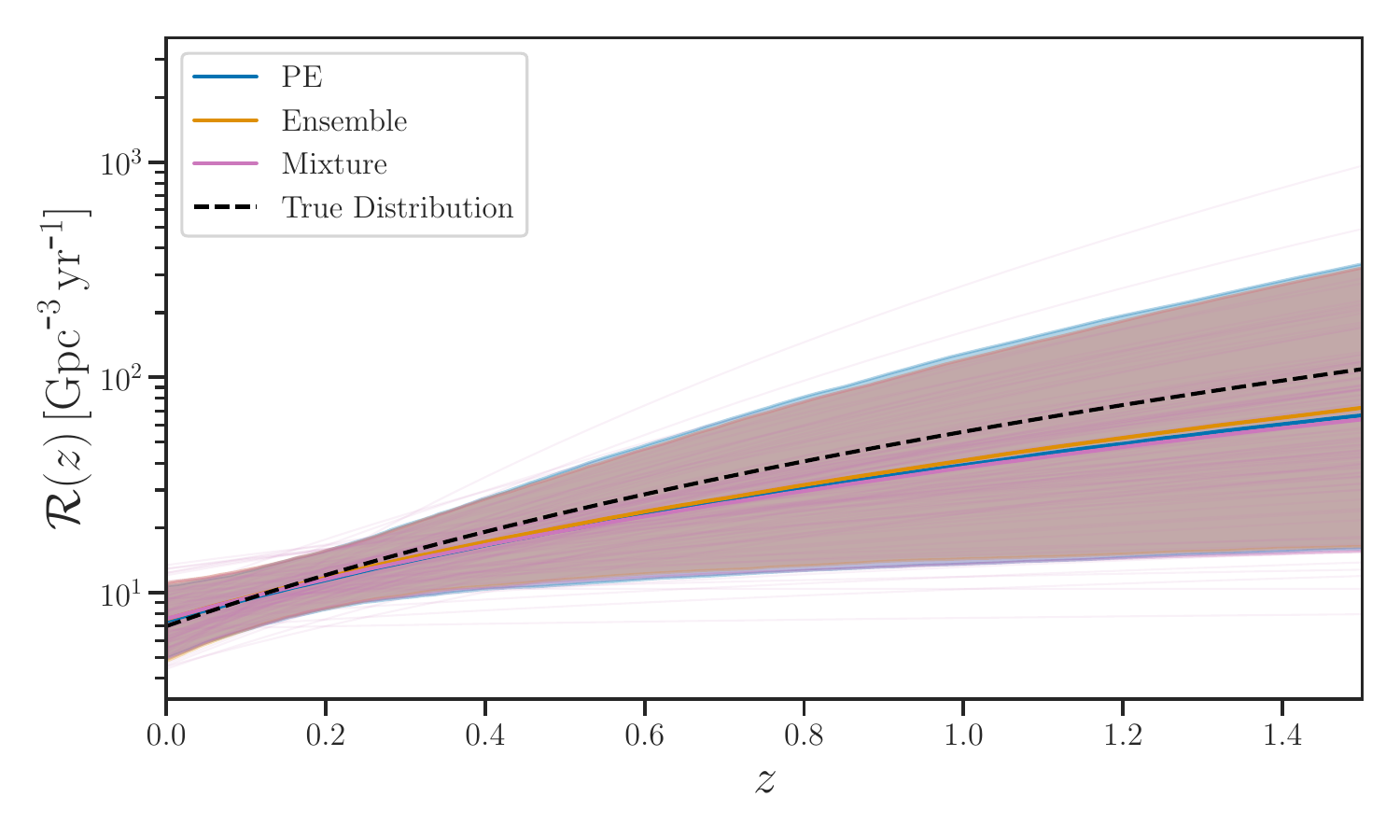}
    \caption{
    Differential merger rates in primary mass $m_1$ (left) and redshift $z$ (right) for the MDC catalog in Fig.~\ref{fig:mdc_pos}.
    The dashed black curves show the injected population; the bands show 90\% credible intervals from PE (blue), Ensemble (orange), and Mixture (magenta).
    The three methods give similar inferred distributions.
    }
    \label{fig:mdc_mass_distributions}
\end{figure*}

We next fix the population at the MDC values given above, including $\lambda=0.04$, and draw 50 catalogs of 69 events from the 276-event pool.
We run hierarchical inference with \texttt{dynesty} \cite{dynesty} for each catalog using PE, Ensemble, and Mixture, with identical priors, sampler settings, and selection samples.
These catalogs can also overlap.
We compare offsets relative to posterior precision using the pull for hyperparameter $\Lambda_a$:
\begin{equation}
    \operatorname{pull}_a=\frac{\overline{\Lambda}_a-\Lambda_{a,\rm true}}{\sigma_{a,\rm posterior}},
\end{equation}
where $\overline{\Lambda}_a$ and $\sigma_{a,\rm posterior}$ are the posterior mean and standard deviation for a given catalog.
Figure~\ref{fig:pull_box} shows the pull distributions across the 50 catalogs.
The gray bands mark offsets of one, two, and three posterior standard deviations; they are not confidence intervals for the ensemble bias.
The three estimators have overlapping pull distributions, with no detectable additional shift from using the GMMs.
The largest offsets occur in the peak location $\mu$ and width $\sigma$, typically at the level of one to two posterior standard deviations.
The small intrinsic peak fraction $\lambda=0.04$ can limit constraints on these parameters, but it does not imply that 4\% of detected events belong to the peak: that fraction also depends on selection effects.
Figure~\ref{fig:mdc_pos} shows a representative catalog for which the three hyperparameter posteriors overlap.

Figure~\ref{fig:mdc_mass_distributions} shows the corresponding primary-mass and redshift merger-rate distributions.
The inferred curves are nearly indistinguishable across the three estimators.
They recover the lower mass cutoff and peak location, while favoring a broader peak than the injected one, consistent with the weak constraints on $\sigma$.
The injected distributions lie within the displayed 90\% credible intervals.

\subsection{Application to GWTC-3}
We apply PE, Ensemble, and Mixture to the GWTC-3 BBH sample \cite{GWTC3} using the public posterior samples and O3 sensitivity inejctions \cite{GWTC3_PE_release, GWTC3_sensitivity_2021}, with the symmetric pairing model in Eqs.~\eqref{eq:pairing} and \eqref{eq:plp_smooth}.
This comparison tests the estimators on observed data with a different population model from the MDC; it is not a test against a known astrophysical truth.

\begin{figure*}[tp]
    \centering
    \includegraphics[width=0.84\textwidth,height=0.70\textheight,keepaspectratio]{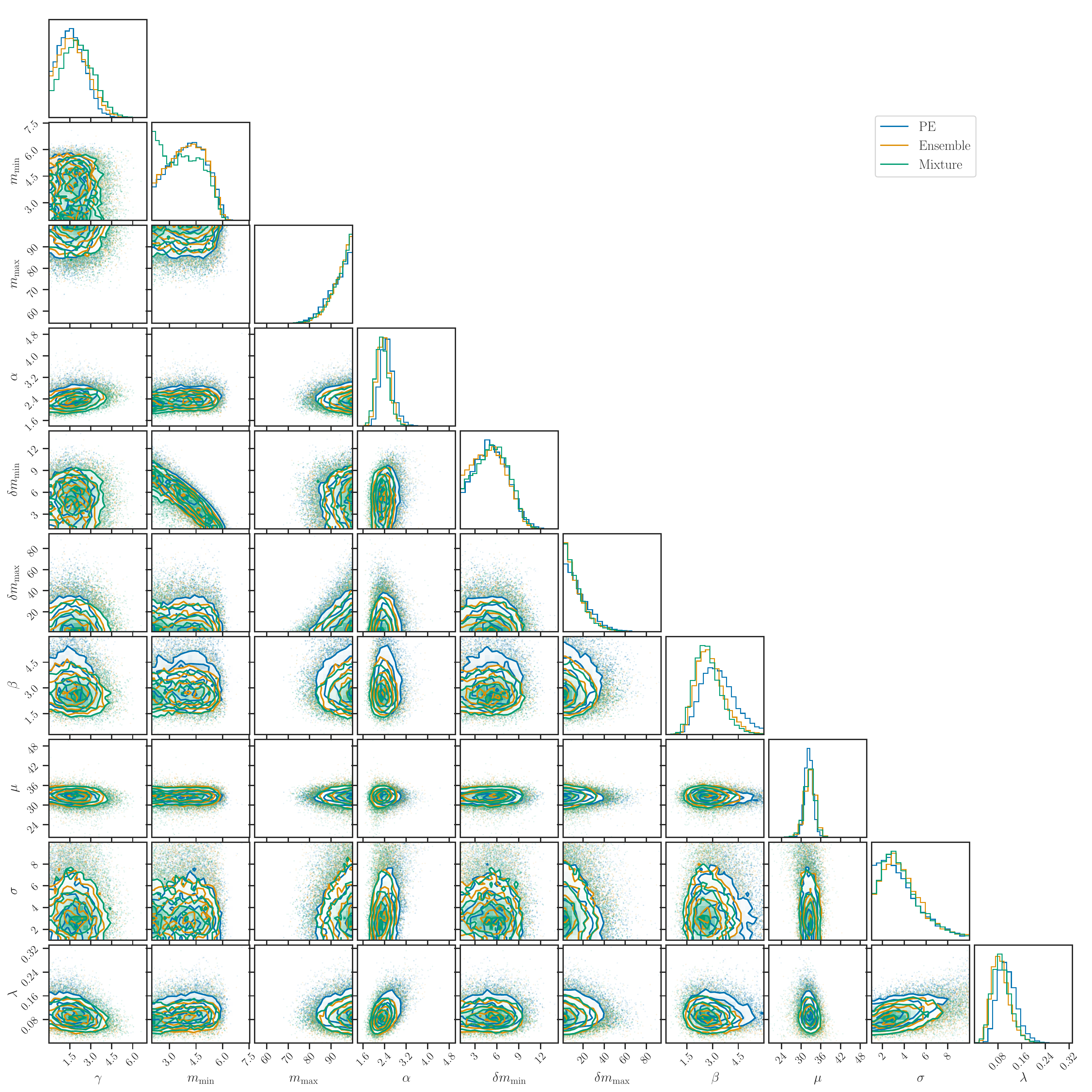}
    \caption{
    Hyperparameter posteriors for the symmetric pairing model fitted to GWTC-3, using PE (blue), Ensemble (orange), and Mixture (green).
    The inferred distributions substantially overlap.
    }
    \label{fig:gwtc3_pos}
\par\medskip
\centering
    \includegraphics[width=0.48\textwidth]{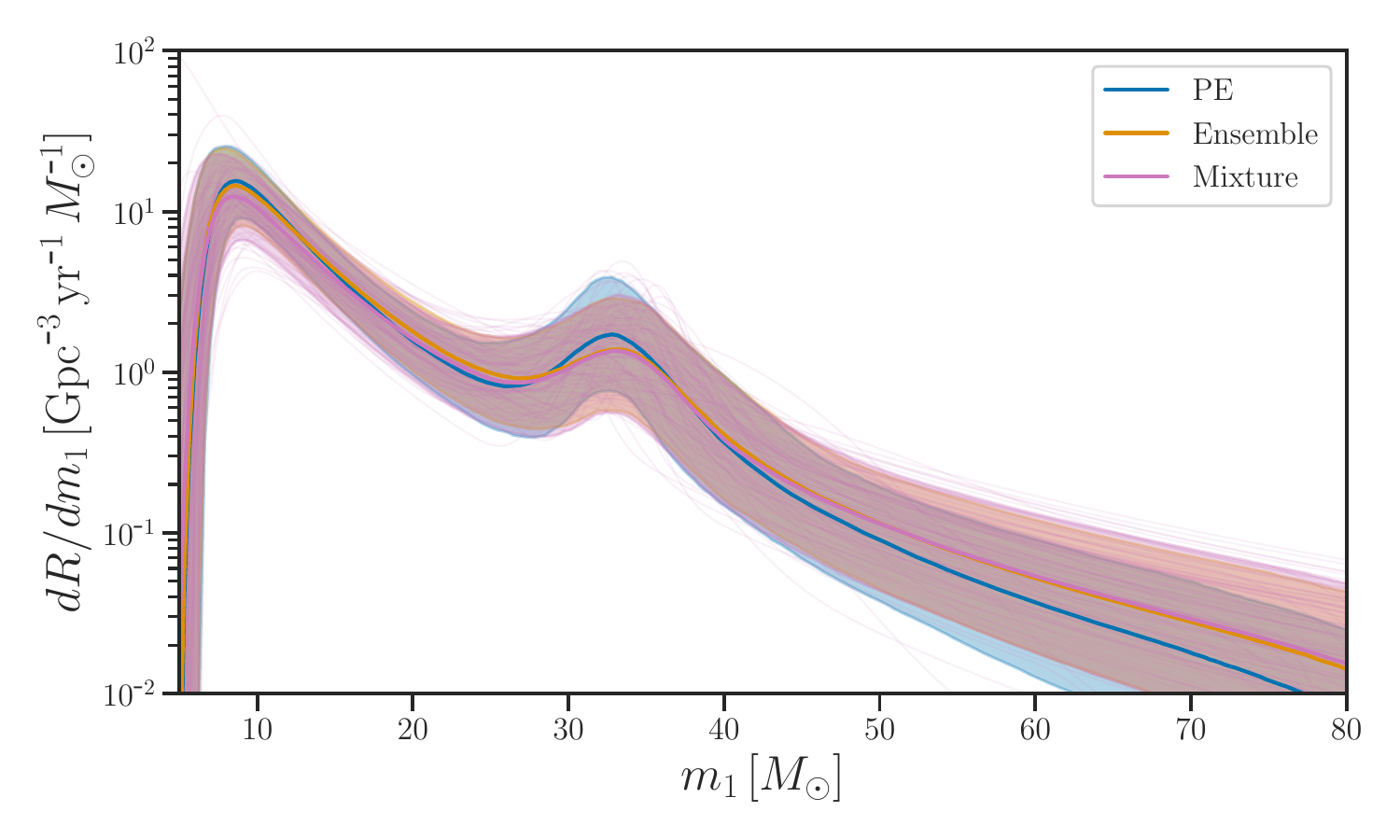}\hfill
    \includegraphics[width=0.48\textwidth]{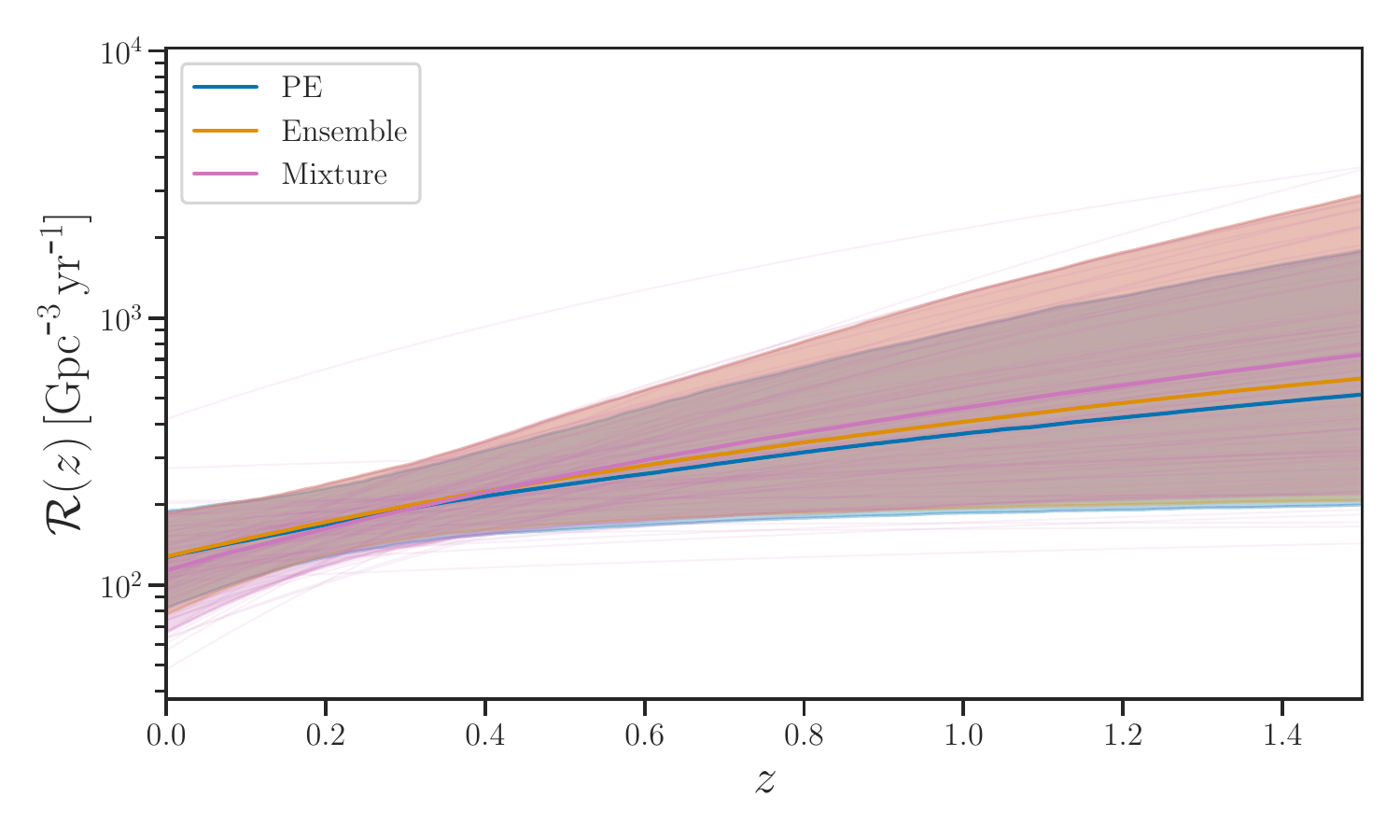}
    \caption{
    Differential merger rates in primary mass $m_1$ (left) and redshift $z$ (right) inferred from GWTC-3.
    The bands show the distributions from PE (blue), Ensemble (orange), and Mixture (magenta), which agree within their posterior uncertainties.
    }
    \label{fig:gwtc3_mass_distributions}
\end{figure*}

Figure~\ref{fig:gwtc3_pos} compares the recovered hyperparameters.
The three estimators give substantially overlapping posteriors and consistent primary-mass and redshift merger-rate distributions (Fig.~\ref{fig:gwtc3_mass_distributions}).

\section{Discussion}\label{sec:discussion}

We fit GMM event posteriors and compare shared ensemble and population-informed importance-sampling proposals.
These separate the Monte Carlo sampling budget from the PE sample count, but not from uncertainty in the fitted densities.

Because the proposals share a sample pool across events, their Monte Carlo errors are correlated.
We include those cross-event covariances in the variance calculation.
Increasing the number of evaluation draws lowers the conditional Monte Carlo variance, but not uncertainty in the fitted densities.
For the 69-event MDC catalogs, the median bootstrap estimate of the event-term fit variance is approximately $0.06$.
This is below the reference total log-likelihood variance of unity, but it is not a complete error budget: selection uncertainty and systematic density-approximation error require separate assessment.
Using all available PE samples rather than a common downsampled set, or averaging multiple density fits, may reduce fit variability without rerunning PE.
Neither possibility was tested here, and the reported value is not a formal upper bound.

The simulated-catalog comparisons show similar parameter recovery for PE, Ensemble, and Mixture.
Their PP curves agree with one another, although the shared departure for $\lambda$, the small number of prior draws, and catalog overlap limit the calibration test.
Their pull distributions also overlap, with the largest offsets in the weakly constrained peak parameters $\mu$ and $\sigma$.
These results show no detectable additional bias from the GMM estimators relative to PE in the tests performed; they do not establish that all three methods are unbiased.

Ensemble does not require an initial population fit, whereas Mixture uses one to select $\Lambda_0$.
For reducing Monte Carlo variance when no population-informed proposal is needed, Ensemble therefore avoids that extra cost.
Mixture may be useful when a fiducial population identifies narrow regions that receive few ensemble draws.
Provided both proposals cover the integrand and the importance weights are correct, however, they estimate the same likelihood, even when the population model is misspecified.
A discrepancy between them should first be investigated as a sampling, density-approximation, or implementation issue; proposal disagreement alone is not a diagnostic of population-model misspecification.

Hussain et al.\ \cite{TGMM} use truncated Gaussian mixtures to mitigate density-estimation bias near physical boundaries, such as zero spin magnitude.
Where the density and population factors have compatible mixture forms, their overlap can be integrated analytically, avoiding Monte Carlo variance for those factors; sampled factors and density-fit uncertainty remain.
Mancarella and Gerosa \cite{Mancarella_2025} adopt an alternative approach, jointly sampling the hyperparameters and the event parameters, with the population reweighting factor entering the target density directly.
This removes the per-event Monte Carlo variance, at a cost of a stricter convergence requirement over the much larger parameter space.
Our estimators retain numerical marginalization over the event parameters, which confines sampling to the lower-dimensional hyperparameter space, and allow correlated population models such as Eq.~\eqref{eq:conditional_plp}.
The trade-off is continued Monte Carlo error, which can be reduced by increasing the proposal sample count.

Adamcewicz et al.\ \cite{flowNPE} also replace discrete event posteriors with continuous densities, using normalizing flows to address the large-catalog sampling problem.
Their posterior-based proposal differs from our population-informed mixture, but both approaches retain uncertainty in the learned density.
GMMs are simple to evaluate, although each full-covariance component has a number of covariance parameters that grows quadratically with dimension.
Normalizing flows offer another representation for higher-dimensional correlated densities \cite{Kobyzev_2021, nf1}; their efficiency depends on architecture and training.
Combining flow-based event densities with the proposals studied here is a possible extension.

Heinzel and Vitale \cite{heinzel2025} quantify information loss from Monte Carlo likelihood approximations and propose a bias correction distinct from the prescription of Farr \cite{Farr_2019}.
Applying that error analysis here requires shared-sample covariances and density-fit uncertainty.
Our tests use two phenomenological mass models, not exact physical relations.
Equation-of-state and cosmological applications therefore need further validation of density accuracy and support.

\section{Acknowledgments}
The authors thank Anarya Ray, Noah E. Wolfe, Sylvia Biscoveanu, Michele Mancarella, and Aditya Vijaykumar for useful discussions.
The authors also gratefully acknowledge the support of LIGO and Virgo for the provision of computational resources.
S.~M.~C.~Y. acknowledges support from the National Science Foundation under grants PHY-2207728 and PHY-2513124.
I.~M.~H. is supported by a McWilliams postdoctoral fellowship at Carnegie Mellon University.
This research used data or software from the Gravitational Wave Open Science Center (\url{https://gwosc.org}), a service of the LIGO Scientific Collaboration, the Virgo Collaboration, and KAGRA.
This material is based upon work supported by NSF's LIGO Laboratory which is a major facility fully funded by the National Science Foundation.

\onecolumngrid
\appendix
\section{Smoothing functions}\label{app:smoothing}
For $x=m-m_{\min}$ and a positive lower smoothing width $\delta m_{\min}$, we define

\begin{equation}
    S_{\rm lo}(m)=
    \begin{cases}
        0, & x\leq0,\\
        [g(x,\delta m_{\min})+1]^{-1}, & 0<x<\delta m_{\min},\\
        1, & x\geq\delta m_{\min},
    \end{cases}
\end{equation}
where
\begin{equation}
    g(x,\delta)=\exp\!\left(\frac{\delta}{x}+\frac{\delta}{x-\delta}\right),
    \qquad 0<x<\delta.
\end{equation}

For the upper cutoff, let $y=m_{\max}-m$ and $\delta m_{\max}>0$. Then
\begin{equation}
    S_{\rm hi}(m)=
    \begin{cases}
        0, & y\leq0,\\
        [g(y,\delta m_{\max})+1]^{-1}, & 0<y<\delta m_{\max},\\
        1, & y\geq\delta m_{\max}.
    \end{cases}
    \label{eq:smax}
\end{equation}
The zero-width limits recover the step functions $S_{\rm lo}(m)=\Theta(m-m_{\min})$ and $S_{\rm hi}(m)=\Theta(m_{\max}-m)$ used in the MDC power-law component; their values exactly at the cutoffs do not affect the continuous density.

\section{Derivation of the estimator variance}\label{app:variance}
For the estimated likelihood in Eq.~\eqref{eq:total_likelihood},
\begin{equation}
    \ln\widehat{\mathcal L}(\Lambda)=
    \sum_{i=1}^{N_{\rm obs}}\ln\widehat{\mathcal L}_i(\Lambda)
    -N_{\rm obs}\ln\hat\xi(\Lambda),
\end{equation}
up to constants independent of $\Lambda$.
Here $\widehat{\mathcal L}(\Lambda)\equiv\widehat{\mathcal L}(\{d_i\}|\Lambda)$.
The event and selection calculations use independent sample pools, so
\begin{equation}
    \Var[\ln\widehat{\mathcal L}]=
    \Var\!\left[\sum_i\ln\widehat{\mathcal L}_i\right]
    +N_{\rm obs}^2\Var[\ln\hat\xi].
\end{equation}

\subsection{Event-likelihood variance}
Let $D$ denote the data used to fit the event densities.
The law of total variance gives
\begin{equation}
\begin{split}
    \Var\!\left[\sum_i\ln\widehat{\mathcal L}_i\right]
    ={}&\underbrace{\mathbb E_D\!\left[\Var_{\rm MC}\!\left(\sum_i\ln\widehat{\mathcal L}_i\mid D\right)\right]}_{\text{conditional Monte Carlo variance}}\\
    &+\underbrace{\Var_D\!\left[\mathbb E_{\rm MC}\!\left(\sum_i\ln\widehat{\mathcal L}_i\mid D\right)\right]}_{\text{density-fit variance}}.
\end{split}
\label{eq:total-var-decomp}
\end{equation}
The second term is a variance of the catalog sum; replacing it with a sum of per-event variances additionally requires independent fit errors.
Conditional on the fitted densities and the proposal,
\begin{align}
    \Var_{\rm MC}\!\left(\sum_i\ln\widehat{\mathcal L}_i\mid D\right)
    &=\sum_i\Var_{\rm MC}[\ln\widehat{\mathcal L}_i\mid D]
      +\sum_{i\ne j}\Cov_{\rm MC}[\ln\widehat{\mathcal L}_i,\ln\widehat{\mathcal L}_j\mid D]\notag\\
    &=\sum_{i,j}C_{ij},
    \label{eq:mc-double-sum}
\end{align}
where $C_{ij}$ is the conditional covariance between the event estimators.
For PE, independently sampled event posteriors give vanishing off-diagonal Monte Carlo covariances.
For Ensemble and Mixture, all event densities are evaluated on the same proposal draws, so these covariances must be retained.

For $\hat L_i\equiv\widehat{\mathcal L}_i=(Z_i/N)\sum_{k=1}^N w_{ik}$ evaluated on a shared pool of independent draws from $q$,
\begin{equation}
\begin{aligned}
    \Cov[\hat L_i,\hat L_j]
    &= \frac{Z_i Z_j}{N^2}\sum_{k,l}\Cov\!\big[w_{ik},w_{jl}\big] \\
    &= \frac{Z_i Z_j}{N^2}\sum_{k}\Cov\!\big[w_{ik},w_{jk}\big] \\
    &= \frac{Z_i Z_j}{N}\big(\langle w_i w_j \rangle - \langle w_i \rangle \langle w_j \rangle\big),
\end{aligned}
\end{equation}
where only the $k=l$ terms survive because the proposal draws are independent. Angle brackets denote expectations over $q$, conditional on the fitted densities.
A first-order delta-method expansion gives
\begin{equation}
    C_{ij}=\Cov[\ln\hat L_i,\ln\hat L_j]
    \simeq\frac{\Cov[\hat L_i,\hat L_j]}{\mathbb E[\hat L_i]\mathbb E[\hat L_j]}
    =\frac{1}{N}\left(\frac{\langle w_iw_j\rangle}
    {\langle w_i\rangle\langle w_j\rangle}-1\right),
\end{equation}
so the evidence factors $Z_i$ cancel.
Define normalized weights $u_{ik}=w_{ik}/\sum_{l=1}^N w_{il}$, so that $\sum_k u_{ik}=1$.
A plug-in estimator of the first-order log covariance is
\begin{equation}
    \widehat C_{ij}=\frac{1}{N-1}\left(N\sum_{k=1}^N u_{ik}u_{jk}-1\right).
\end{equation}
The factor $N-1$ is the Bessel correction for the sample covariance of the weights; it does not make the nonlinear log-covariance approximation exactly unbiased.
Summing over all pairs of events gives
\begin{align}
    \widehat{\Var}_{\rm MC}\!\left(\sum_i\ln\hat L_i\mid D\right)
    &=\sum_{i,j}\widehat C_{ij}\notag\\
    &=\frac{1}{N-1}\left[N\sum_k\left(\sum_i u_{ik}\right)
      \left(\sum_j u_{jk}\right)-\sum_{i,j}1\right]\notag\\
    &=\frac{1}{N-1}\left[N\sum_{k=1}^N\left(\sum_i u_{ik}\right)^2-N_{\rm obs}^2\right].
    \label{eq:var-total-perevent}
\end{align}

The density-fit term in Eq.~\eqref{eq:total-var-decomp} arises because each GMM is trained on a finite PE sample set.
We estimate its per-event contribution by drawing $B=50$ bootstrap resamples, each of size 5000, with replacement from the event's $N_{\rm PE}=5000$ samples.
We fit a GMM $\hat p_i^{(b)}$ to each resample and evaluate its likelihood using a large, fixed pool of draws from $q$.
Keeping this pool fixed reduces Monte Carlo differences between refits; a sufficiently large pool is still needed to isolate fit variability.
The bootstrap estimator for one event is
\begin{equation}
    \widehat{\Var}_{\rm fit}[\ln\hat L_i] =
    \frac{1}{B-1} 
    \sum_{b=1}^B \left( \ln \hat{\mathcal{L}}_i^{(b)} - \overline{\ln \hat{\mathcal{L}}_i} \right)^2,
    \label{eq:fit_var}
\end{equation}
where the mean bootstrap log-likelihood is
\begin{equation}
    \overline{\ln \hat{\mathcal{L}}_i} = \frac{1}{B} \sum_{b=1}^B \ln \hat{\mathcal{L}}_i^{(b)}.
\end{equation}
We sum these per-event estimates under the approximation that independently fitted event densities have independent fit errors and that residual Monte Carlo error is negligible.
Shared-proposal effects at finite evaluation count, uncertainty in the fitted reference population, and systematic density-approximation bias are not measured by this sum.

\subsection{Selection-efficiency variance}
To first order the log-selection variance is
\begin{equation}
    \Var[\ln\hat\xi]\simeq\frac{\Var[\hat\xi]}{\xi^2}\equiv\frac{\sigma_\xi^2}{\xi^2}.
\end{equation}
For a Monte Carlo estimator $\hat I=\tfrac1M \sum_{j=1}^{M} f(\theta_j)/p(\theta_j)$
built from $M$ independent draws $\theta_j\sim p(\theta)$, the unbiased variance
estimator is
\begin{equation}
  \hat\sigma^2_{I}
  = \frac{1}{M-1}\left(
      \frac{1}{M}\sum_{j=1}^{M}
      \frac{f(\theta_j)^2}{p(\theta_j)^2}
      - \hat I^2 \right).
  \label{eq:var-mc-generic}
\end{equation}

Applying Eq.~\eqref{eq:var-mc-generic} to the direct-injection estimator in Eq.~\eqref{eq:standard_selection} gives

\begin{equation}
  \hat\sigma^2_{\xi,\rm inj}
  =\frac{1}{N_{\rm inj}-1}\left[
      \frac{1}{N_{\rm inj}}\sum_{j=1}^{N_{\rm found}}
      \frac{p_{\rm pop}(\vartheta_j|\Lambda)^2}
      {p_{\rm inj}(\vartheta_j|\Lambda_{\rm inj})^2}
      -\hat\xi_{\rm inj}^2\right],
  \label{eq:var-inj-est}
\end{equation}

The sum runs over the $N_{\rm found}$ detected injections; undetected injections have zero weight. For $N_{\rm GMM}$ independent draws from the fitted approximation to $q_{\rm sel}$ in Eq.~\eqref{eq:sel_proposal}, the estimator is

\begin{equation}
  \hat\xi_{\rm GMM}(\Lambda)
  =\frac{\eta}{N_{\rm GMM}}\sum_{j=1}^{N_{\rm GMM}}
      \frac{p_{\rm pop}(\vartheta_j|\Lambda)}
      {p_{\rm inj}(\vartheta_j|\Lambda_{\rm inj})},
  \label{eq:xi-gmm-est}
\end{equation}

\begin{equation}
  \hat\sigma^2_{\xi,\rm GMM}
  =\frac{1}{N_{\rm GMM}-1}\left[
      \frac{\eta^2}{N_{\rm GMM}}\sum_{j=1}^{N_{\rm GMM}}
      \frac{p_{\rm pop}(\vartheta_j|\Lambda)^2}
      {p_{\rm inj}(\vartheta_j|\Lambda_{\rm inj})^2}
      -\hat\xi_{\rm GMM}^2\right].
  \label{eq:var-gmm-est}
\end{equation}

This is the Monte Carlo variance conditional on the fitted injection density and a fixed estimate of $\eta$; it excludes uncertainty in those quantities.
In either case, the relative-variance estimator can be written as
\begin{equation}
    \frac{\hat\sigma_\xi^2}{\hat\xi^2}
    =\frac{1}{N-1}\left(\frac{N}{N_{{\rm eff},\xi}}-1\right),
    \qquad
    N_{{\rm eff},\xi}=\frac{(\sum_j w_j)^2}{\sum_j w_j^2}.
    \label{eq:relvar-Neff}
\end{equation}
For direct injections, $N=N_{\rm inj}$ and the weights include the detection indicators.
For GMM draws, $N=N_{\rm GMM}$ and the weights include $\eta$.
The constant $\eta$ cancels from the conditional relative variance, but its estimation uncertainty does not thereby disappear from the full selection error budget.

\clearpage
\section{GMM fit diagnostics}\label{app:fit}

\begin{figure}[h]
    \centering
    \includegraphics[width=0.57\linewidth]{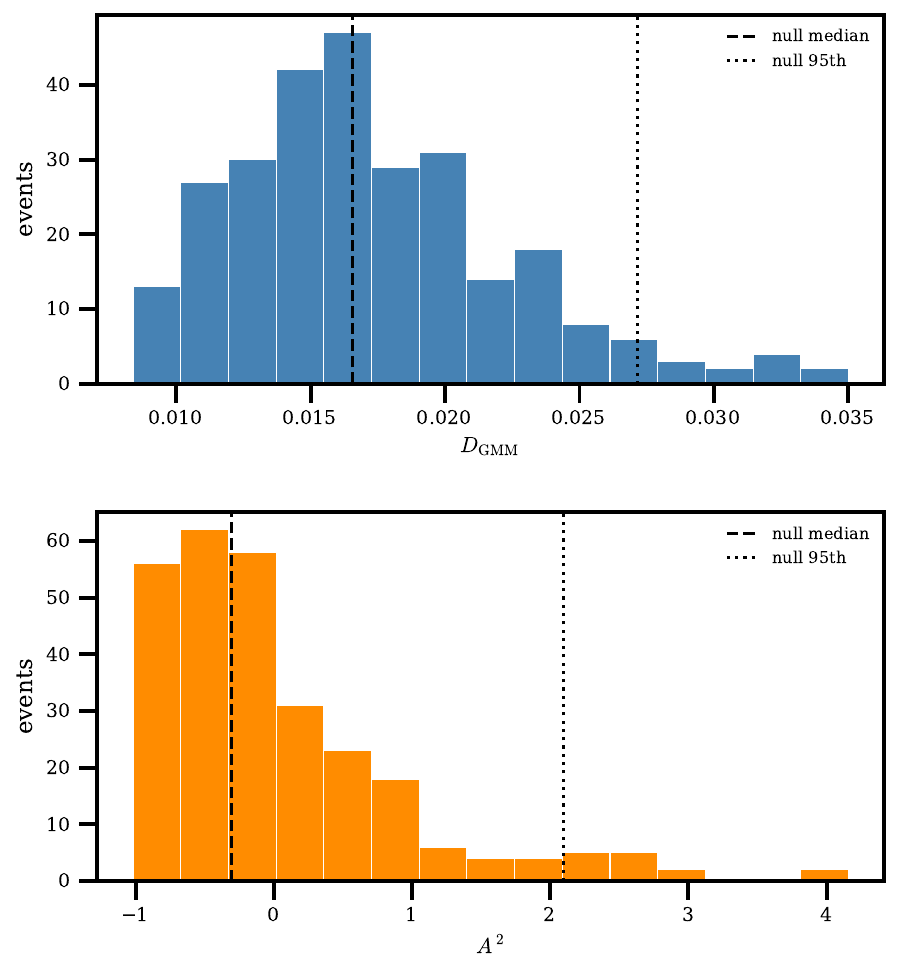}
    \caption{
    Two-sample Kolmogorov--Smirnov (KS) statistic $D_{\rm GMM}$ and Anderson--Darling (AD) statistic $A^2$ for detector-frame primary mass across the 276 simulated events.
    Dashed lines mark the reported reference medians and 95th percentiles.
    These are marginal, in-sample diagnostics: the comparison PE samples were also used to fit the GMMs.
    }
    \label{fig:KS_AD_tests}
\end{figure}

We compare samples drawn from each fitted GMM with the corresponding PE samples using two-sample Kolmogorov--Smirnov (KS) and Anderson--Darling (AD) statistics.
The AD statistic gives greater weight to tail differences than the KS statistic.
For independent samples from the same continuous distribution, the appropriately scaled KS statistic has the asymptotic Kolmogorov distribution.
For the AD reference distribution, we draw both comparison sets from the fitted GMMs.
The reference medians and 95th percentiles are shown in Fig.~\ref{fig:KS_AD_tests}.

For detector-frame primary mass, 4.7\% of the 276 events exceed the reported KS 95th-percentile threshold and 5.4\% exceed the corresponding AD threshold.
These fractions suggest useful marginal agreement, but they are not a calibrated test of the full density fit: the PE samples used for comparison also trained the GMMs.
An independent held-out comparison, or a null procedure that repeats the fitting step, is needed to calibrate the diagnostic after fitting.
Agreement in this one-dimensional marginal also does not establish accuracy of the joint density or of sparsely sampled tails.

\clearpage
\twocolumngrid
\interlinepenalty=10000
\bibliographystyle{apsrev4-2}
\bibliography{refs}
\end{document}